\documentclass[12pt]{article}
\pdfoutput=1
\usepackage{yfonts}
\usepackage{color}
\usepackage{mhchem}
\usepackage{xcolor}
\usepackage{cite}
\usepackage{hyperref}
\hypersetup{colorlinks=true,linkcolor=red,anchorcolor=black,citecolor=green}
\usepackage[toc,page]{appendix}
\usepackage{amsfonts}
\usepackage{bbold}
\usepackage{textcomp}
\usepackage[DIV13]{typearea}
\usepackage{amsmath, amsthm, amssymb, mathtools,empheq,latexsym,dsfont}
\usepackage{bbm}
\usepackage{slashed, simplewick}
\usepackage[utf8]{inputenc}
\usepackage{graphicx,placeins}
\usepackage{makeidx}
\usepackage[font=small,labelfont=bf]{caption}
\usepackage{nicefrac}
\usepackage{subfigure}
\usepackage{array, bigdelim,multirow,multicol}
\usepackage[integrals]{wasysym}
\usepackage{fancybox}
\usepackage{bm}
\usepackage{float}
\usepackage{rotating}
\usepackage{colortbl}
\usepackage{makecell}
\usepackage{pifont}
\usepackage{longtable}
\usepackage{booktabs}
\usepackage[top=2cm,textwidth=16.6cm,textheight=22.75cm]{geometry}
\usepackage{doi}
\usepackage{fancybox}
\usepackage[compat=1.0.0]{tikz-feynman}
\usepackage{tikz}
\usepackage{empheq}
\allowdisplaybreaks
\newcommand{\ignore}[1]{}
\usepackage{enumitem}
\usepackage{gensymb}

\usepackage{braket}

\graphicspath{{immagini/}}
\definecolor{Gray}{gray}{0.92}
\DeclareMathAlphabet{\mathscr}{OT1}{pzc}{m}{it}

\newcommand{\be}{\begin{equation}}
\newcommand{\ee}{\end{equation}}
\newcommand{\bea}{\begin{eqnarray}}
\newcommand{\eea}{\end{eqnarray}}

\makeatletter
\@addtoreset{equation}{section}
\makeatother

\begin{document}

\unitlength = 1mm
\setlength{\extrarowheight}{0.2 cm}
\thispagestyle{empty}
\bigskip
\vskip 1cm

\title{\Large \bf Texture-zeros in minimal seesaw from non-invertible symmetry fusion rules \\[2mm] }

\date{}

\author{Zheng Jiang$^{a}$\footnote{E-mail: {\tt
jzdyx@mail.ustc.edu.cn}},\
Bu-Yao Qu$^{a}$\footnote{E-mail: {\tt
qubuyao@mail.ustc.edu.cn}},  \
Gui-Jun Ding$^{a,b}$\footnote{E-mail: {\tt
dinggj@ustc.edu.cn}}
\\*[20pt]
\centerline{
\begin{minipage}{\linewidth}
\begin{center}
$^a${\it \small Department of Modern Physics,  and Anhui Center for fundamental sciences in theoretical physics,\\
University of Science and Technology of China, Hefei, Anhui 230026, China}\\[2mm]
$^b${\it \small College of Physics, Guizhou University, Guiyang 550025, China}
\end{center}
\end{minipage}}
\\[10mm]}
\maketitle
\thispagestyle{empty}

\centerline{\large\bf Abstract}

\begin{quote}
\indent

The $Z_2$ gauging of $Z_N$ symmetry can enforce certain elements of the fermion Yukawa couplings to vanish. We have performed a systematical study of texture zero patterns of lepton mass matrices in the  minimal seesaw model, and we present all the possible patterns of the charged lepton Yukawa coupling $Y_E$, neutrino Yukawa coupling $Y_{\nu}$, right-handed neutrino mass matrix $M_R$ and the light neutrino mass matrix $M_{\nu}$ which can be derived from the $Z_2$ gauging of $Z_N$ symmetry. The realization of the textures with the maximum number of zeros and the second maximum number of zeros from non-invertible symmetry is studied, and the phenomenological implications in neutrino oscillation are discussed.

\end{quote}

\clearpage

\section{Introduction}
It is well-known that the masses and flavor mixing of quarks and leptons exhibit a peculiar hierarchical pattern~\cite{ParticleDataGroup:2024cfk}. The mass spectrum spans more than twelve orders of magnitude, from sub-eV neutrinos to the heavy top quark with a mass about 173 GeV. The quark flavor mixing is described by the Cabibbo-Kobayashi-Maskawa (CKM) matrix which is parameterized by three mixing angles $\theta^{q}$, $\theta^q_{13}$, $\theta^q_{23}$ and one CP-violating phase $\delta^q_{CP}$. It is known experimentally that the three quark angles are small and hierarchical with $\theta^q_{12}\sim\lambda$, $\theta^q_{23}\sim\lambda^2$, $\theta^q_{13}\sim\lambda^3$, where $\lambda\approx0.225$. In analogy the lepton flavor mixing is described by Pontecorvo-Maki-Nakagawa-Sakata (PMNS) matrix which are parameterized by three lepton mixing angles $\theta^{\ell}_{12}$, $\theta^{\ell}_{13}$, $\theta^{\ell}_{23}$, one Dirac CP violation phase $\delta^{\ell}_{CP}$ and two additional Majorana CP phases $\alpha_{21}$ and $\alpha_{31}$ if neutrinos are their own antiparticles~\cite{ParticleDataGroup:2024cfk}.  Neutrino oscillation experiments reveal two large mixing angles $\theta^{\ell}_{12}\approx33.4^{\circ}$, $\theta^{\ell}_{23}\approx43.5^{\circ}$ and one relatively small angle $\theta^{\ell}_{13}\approx8.6^{\circ}$~\cite{Capozzi:2025wyn}. Similar results are reached in other global analysis of neutrino oscillation data~\cite{Esteban:2024eli,deSalas:2020pgw}. Understanding the above patterns of masses and mixing angles of the quarks and leptons, along with the sources and magnitude of CP violation, is still a big challenge. The fundamental principle behind the structure of fermion masses and mixing is still unknown, see Refs.~\cite{Feruglio:2019ybq,Xing:2020ijf,Ding:2023htn,Ding:2024ozt}  for recent review and Ref.~\cite{Feruglio:2025ztj} for a pedagogic introduction on this topic.

Within the Standard Model (SM), the fermion masses and mixing arise from the Yukawa couplings which are free parameters. In the absence of a complete flavor theory, a popular phenomenological approach is to assume specific texture of quark and lepton mass matrices. The simplest attempt is to assume some entries of the fermion mass matrix vanishing in a specific weak-interaction basis~\cite{Weinberg:1977hb,Fritzsch:1977za}, it is called texture zero in the literature. The assumption of texture zero reduces the number of free parameters, leading to testable relations among fermion masses, mixing angles, and CP-violating phases. For instance, the Fritzsch six-zero ansatz for the quark mass matrices can give rise to the famous relation $\sin\theta_{c}\approx\sqrt{m_d/m_s}$, which elegantly related  mass ratio of down-quark and strange quark to the Cabibbo angle $\theta_{\rm C}$. Likewise the texture zero of neutrino mass matrix in the charged lepton diagonal basis has been extensively studied~~\cite{Frampton:2002yf,Xing:2002ta}. The vanishing entries yield enough constraints so that ones can not only explain the neutrino oscillation data but also make predictions for yet-undetermined parameters such as the absolute neutrino mass scale, the ordering of neutrino masses or the lepton CP violating phases~\cite{Fritzsch:1999ee,Gupta:2012fsl}.

From a theoretical standpoint, the texture zeros may originate from the spontaneous or explicit breaking of
Abelian~\cite{Grimus:2004hf,GonzalezFelipe:2014zjk} or non-Abelian flavor symmetries~\cite{Zhang:2019ngf,Lu:2019vgm,Kikuchi:2022svo,Ding:2022aoe}, or from the structure of higher-dimensional operators in effective field theories. The texture zero may reflect some underlying selection rules. In the last several years, the concept of symmetry was generalized~\cite{Gaiotto:2014kfa} and the applications of non-invertible symmetries to particle physics have been extensively discussed, see Refs.~\cite{Cordova:2022ruw,Schafer-Nameki:2023jdn,Brennan:2023mmt,Bhardwaj:2023kri,Shao:2023gho} for recent review on non-invertible symmetries. It was found that the fusion rules of the non-invertible symmetry can naturally give rise to texture zeros in Yukawa couplings ~\cite{Kobayashi:2024cvp,Kobayashi:2025znw,Kobayashi:2025ldi}. The quark Yukawa textures from non-invertible symmetry has been explored in~\cite{Kobayashi:2024cvp,Kobayashi:2025znw}. It is notable that one can use non-invertible symmetry allows to construct certain texture-zero pattern for quark mass matrices, which can resolve the strong CP problem without invoking the axion~\cite{Liang:2025dkm,Kobayashi:2025thd}. The possible textures of neutrino and charged lepton mass matrices are studied under the assumption that the light neutrino masses are described by the Weinberg operator~\cite{Kobayashi:2025ldi}.

The discovery of neutrino oscillations shows that neutrinos have masses. To account for the tiny neutrino masses, the most straightforward and elegant way is to extend the SM by introducing right-handed neutrinos which are singlets of SM. Although the number of right-handed neutrinos is arbitrary in principle, at least two are necessary to explain the two mass-squared differences observed by solar and atmospheric neutrino oscillation experiments. This is the so-called minimal seesaw model~\cite{King:1999mb,Frampton:2002qc} in which the lightest neutrino is massless, namely $m_1=0$ for normal neutrino mass ordering (NO) and $m_3=0$ for inverted neutrino mass ordering (IO). Although the minimal seesaw model can accommodate tiny neutrino masses by extending the SM at the least cost, it can not predict neither the lepton mixing parameters nor the neutrino mass spectrum. The number of free parameters in the minimal seesaw model is still larger than the number of lower energy observables. One way of reducing the number of free parameters and increasing the predictive power is to assume texture zero in the lepton Yukawa couplings and mass matrix of right-handed neutrinos, which could possibly be realized by imposing $U(1)$ or $Z_n$ Abelian flavor symmetry~\cite{Kageyama:2002zw,Grimus:2004hf,Barreiros:2018ndn}.

In the present work, by using $Z_2$ gauging of $Z_N$ symmetries,  we shall study the patterns of the texture zeros of the lepton Yukaka matrices and the right-handed neutrino mass matrix in the framework of minimal seesaw model. It is notable that there are some texture zero patterns which can not be derived from the conventional Abelian flavor symmetry. The corresponding phenomenological predictions for lepton mixing parameters are studied.

The remaining part of this paper is organized as follows. We recapitulate the formalism of the $Z_2$ gauging of $Z_N$ symmetries in section~\ref{sec:non-invertible-rule}, and the non-invertible selection rules are presented. In section~\ref{sec:texture-zeros-MSM}, we perform a comprehensive study of texture zero of charged lepton mass matrix, Dirac neutrino mass matrix and right-handed neutrino mass matrix which can be obtained from $Z_2$ gauging of $Z_N$ symmetries with $N=3,4,5$. Furthermore, we identify the viable textures with the maximum number of zeros and the second maximum number of zeros in section~\ref{sec:pheno}, and the phenomenological implications in neutrino oscillation experiments are discussed. The new textures of the charged lepton and neutrino Yukawa couplings, which can be generated from the non-invertible $Z_N$ symmetry for $N\geq5$, are discussed. Finally we summarize our main conclusion in section~\ref{sec:conclusion}. We list the possible textures of the neutrino Yukawa coupling $Y_{\nu}$, right-handed neutrino mass matrix $M_R$ and the corresponding light neutrino mass matrix $M_{\nu}$ which can be derived from the $Z_2$ gauging of $Z_N$ symmetry for $N=4, 5$ in Appendix~\ref{app:neutrino-textures-N45}.  Similarly the $Z_2$ gauging of $Z_4$ and $Z_5$ symmetries restricts the allowed forms of the charged lepton Yukawa coupling $Y_E$ which are presented in Appendix~\ref{app:yetex-product}. The texture zeros of the Yukawa couplings $Y_E$ and $Y_{\nu}$ derived from $Z_2$ gauging of $Z_N$ symmetries for $N=6, 7$ are collected in Appendix~\ref{app:higher-N-texproduct}.

\section{\label{sec:non-invertible-rule} Non-invertible selection rules}

In ordinary group-theoretic symmetries, the multiplication law takes the form
\begin{eqnarray}
ab = c
\end{eqnarray}
where $a$, $b$ and $c$ are elements of a group $G$. In the case that $G$ is a Abelian group, if a field $\phi_a$ transforms in the representation corresponding to $a$, and similarly for $\phi_b$ and $\phi_c$, then the process $\phi_a + \phi_b \rightarrow \phi_c$ is allowed when $ab=c$. Otherwise, it is forbidden. Different fields may correspond to the same group element, in which case the same selection rule applies. This yields unique and deterministic coupling constraints since the right-hand side of a product is fixed.

Non-invertible symmetries generalize this notion by replacing the unique group multiplication with fusion rules of the form
\begin{eqnarray}
a b = \sum_{c} N_{ab}^c c
\end{eqnarray}
where $a$, $b$ and $c$ are symmetry operators and $N_{ab}^c \in \mathbb{Z}_{\geq 0}$ denoted fusion coefficients. A process $\phi_a + \phi_b \rightarrow \phi_c$ is allowed if $N_{ab}^c\neq 0$, and forbidden otherwise.
Crucially, the product $ab$ need not yield a single output. The right-hand side may contain multiple operators, reflecting the non-invertibility of the symmetry. The richer algebraic structure leads to selection rules that differ qualitatively from those of ordinary group theory.

A concrete and physically motivated realization of such fusion rules in four dimensional quantum field theories arises from gauging outer automorphisms of a symmetry group~\cite{Bhardwaj:2022yxj,Bartsch:2022mpm}.
In particular, explicit models can be constructed by performing the simplest $Z_2$ gauging of a $Z_N$ symmetry~\cite{Kobayashi:2024yqq,Kobayashi:2024cvp}.
Let $g$ be a generator of $Z_N$, so that $g^N = \mathbb{1}$. We consider the $Z_2$ outer automorphism
\begin{eqnarray}
e g^k e^{-1} = g^k\,,~~~~ r g^k r^{-1} = g^{-k}\,.
\end{eqnarray}
Using this automorphism, one defines equivalence classes
\begin{eqnarray}
[g^k] = \{ g^k, g^{-k} \}~~~~ k = 0, 1, \cdots \, \left\lfloor \dfrac{N}{2} \right\rfloor \,,
\end{eqnarray}
where $\lfloor x \rfloor$ denotes the floor function, i.e., the greatest integer less than or equal to $x$.
In the following, we use the $[k]$ as a shorthand notation for the conjugacy class $[g^k]$, for simplicity. The fusion rules among these classes take the form
\begin{eqnarray}
\label{eq:fusion-rule-zngagugez2} [k][k'] = [k+k'] + [k-k'] \,.
\end{eqnarray}
Which is precisely the algebra obtained by $Z_2$ gauging of $Z_N$.
It can be observed that the group $Z_N$ with its outer automorphisms $\{e,r\}$ is isomorphic to the dihedral group $D_M \simeq Z_N \rtimes \mathbb{Z}_2$.
The above fusion rules form a subset of the fusion rules among the conjugacy classes of the dihedral group $D_M$~\cite{Dong:2025jra}.
Such $Z_2$ gauging of $Z_N$ can be engineered in higher-dimensional field theories in string compactifications. For instance, magnetized compactifications naturally yield $Z_N$ flavor symmetries.

Coupling selection rules now follow directly from the fusion algebra. A two point coupling $\phi_{k_1}\phi_{k_2}$ is allowed if $[0]$ appears on the right-hand side of the fusion of $[k_1]$ and $[k_2]$, which occurs when
\begin{eqnarray}
[0] \in \{ [k_1+k_2], [k_1-k_2] \}
\end{eqnarray}
which is equivalent to the condition
\begin{eqnarray}
k_1 \pm k_2 = 0 ~~({\rm mod}~N) \,.
\end{eqnarray}
This implies, in particular, that two-point couplings, including mass terms and kinetic terms, are permitted only between fields in the same class. For the three-point couplings, a term of the form $\phi_{k_1} \phi_{k_2} \phi_{k_3}$ is allowed if
\begin{eqnarray}
[0] \in \{ [k_1+k_2+k_3], [k_1+k_2-k_3], [k_1-k_2+k_3], [k_1-k_2-k_3] \}
\end{eqnarray}
which is equivalent to
\begin{eqnarray}
k_1 \pm k_2 \pm k_3 = 0 ~~({\rm mod}~M) \,.
\end{eqnarray}
More generally, an $n$-point coupling $\phi_{k_1} \cdots \phi_{k_n}$ is allowed if
\begin{eqnarray}
\sum_{i=1}^{n} \pm k_i= 0 ~~({\rm mod}~M) \,.
\end{eqnarray}
These generalized selection rules reflect the non-invertible nature of the symmetry. They are not simply additive charge conservation laws, but are instead dictated by multi-valued fusion structure originating from $Z_2$ gauging.

\section{\label{sec:texture-zeros-MSM} Minimal seesaw model and texture zeros of lepton mass matrices from non-invertible selection rules }

In the present work, we focus on the minimal seesaw model featuring only two generations of right-handed neutrinos denoted as $\nu_{R1}$ and $\nu_{R2}$~\cite{King:1999mb,Frampton:2002qc}, see Ref.~\cite{Xing:2020ald} for a review. The gauge invariant Lagrangian for the charged lepton and neutrino masses can be written as
\begin{equation}
\mathcal{L}_{\text{lepton}} =-(Y_{E})_{ij} \overline{\ell_{Li}} H E_{Rj}-(Y_{\nu})_{ij} \overline{\ell_{Li}} \widetilde{H} \nu_{Rj}-\frac{1}{2}(M_{R})_{ij}\overline{\nu_{Ri}^C}  \nu_{Rj} \,,
\end{equation}
where $i,j=1, 2, 3$ are the flavor indices, $\ell_{L}$ and $H$ stand for the SM left-handed lepton doublets and Higgs doublet respectively, $E_R$ denotes the right-handed charged lepton fields. We have defined $\widetilde{H}=i\sigma_2 H^*$, with $\sigma_2$ the second Pauli matrix, and $\nu_{Ri}^C= \mathcal{C}\overline{\nu_{Ri}}^T$ with $\mathcal{C}=i\gamma^{0}\gamma^2$ being the charge conjugation matrix. Moreover, the charged lepton Yukawa coupling matrix $Y_E$ and the neutrino Yukawa coupling matrix $Y_{\nu}$ are $3\times3$ and $3\times 2$ general complex matrix, and the right-handed neutrino mass $M_R$ is $2\times 2$ symmetric and complex matrix.

After the electroweak gauge symmetry spontaneous breaking by the vacuum expectation value of the Higgs field $\langle H\rangle=\left(0, v/\sqrt{2}\right)^T$ with $v\simeq 246$ GeV, the charged lepton and Dirac neutrino mass matrices are obtained as
\begin{equation}
M_E=\frac{1}{\sqrt{2}}Y_E v\,,~~~M_D=\frac{1}{\sqrt{2}}Y_{\nu} v\,.
\end{equation}
In the limit $M_R\gg v$, integrating out the heavy right-handed neutrinos, the effective light neutrino mass matrix is given by the seesaw formula
\begin{equation}
M_\nu=-M_DM^{-1}_RM^T_D=-\frac{1}{2}v^2Y_{\nu}M_R^{-1}Y_{\nu}^T \,.\label{eq:eff-neu-mass}
\end{equation}
Both $M_{E}$ and $M_{\nu}$ can be diagonalized by bi-unitary transformations,
\begin{eqnarray}
 U_{\ell_L}^\dagger M_E U_{E_R}=\begin{pmatrix}
m_e ~&~ 0 ~&~ 0 \\
0 ~&~ m_\mu ~&~ 0 \\
0 ~&~ 0 ~&~ m_\tau
\end{pmatrix}\,,~~~
U_{\nu_L}^{\dagger}M_\nu U^{*}_{\nu_L}=\begin{pmatrix}
 m_1 ~&~ 0 ~&~ 0 \\
 0 ~&~ m_2 ~&~ 0 \\
 0 ~&~ 0 ~&~ m_3
\end{pmatrix} \label{eq:UnuL}\,.
\end{eqnarray}
The mismatch between the flavor eigenstates and mass eigenstates results in the lepton mixing matrix in the charged current interaction, i.e.
\begin{equation}
U=U_{\ell_L}^\dagger U_{\nu_L}\,.
\end{equation}
Here $U$ is the so-called Pontecorvo-Maki-Nakagawa-Sakata (PMNS) matrix, and it is parameterized as~\cite{ParticleDataGroup:2024cfk}
\begin{equation}
\label{eq:PMNS-para}U=\begin{pmatrix}
c_{12}c_{13} & s_{12}c_{13} & s_{13}e^{-i\delta_{CP}} \\
-s_{12}c_{23}-c_{12}s_{23}s_{13}e^{i\delta_{CP}} & c_{12}c_{23}-s_{12}s_{23}s_{13}e^{i\delta_{CP}} & s_{23}c_{13} \\
s_{12}s_{23}-c_{12}c_{23}s_{13}e^{i\delta_{CP}} & -c_{12}s_{23}-s_{12}c_{23}s_{13}e^{i\delta_{CP}} & c_{23}c_{13}
\end{pmatrix}
\begin{pmatrix}
 1 & 0 & 0 \\
 0 & e^{i\alpha_{21}/2} & 0 \\
 0 & 0 & e^{i\alpha_{31}/2}
\end{pmatrix} \,,
\end{equation}
where $c_{ij}=\cos\theta_{ij}$, $s_{ij}=\sin\theta_{ij}$,  $\theta_{ij}\in[0, \pi/2)$ are the three lepton mixing angles, $\delta_{CP}\in[0, 2\pi)$ is the Dirac CP
violation phase, and $\alpha_{21}$, $\alpha_{31}$ are two Majorana CP violation phases. In the minimal seesaw model, the lightest neutrino is massless so that we have $m_1=0$ for normal ordering (NO) neutrino mass spectrum and $m_3=0$ for inverted ordering (IO) neutrino mass spectrum~\cite{King:1999mb,Frampton:2002qc}. As a consequence, $\alpha_{21}$ is the unique physical Majorana CP phase and $\alpha_{31}$ can be absorbed by field redefinition.

In the following, we study which texture zero patterns of the Yukawa matrices $Y_{E}$, $Y_{\nu}$ and the mass matrix $M_{R}$ can be derived from the $Z_2$ gauging of $Z_N$ symmetries with $N=3,4,5$. The three generations of lepton fields $\ell_{Li}$, $E_{Ri}\;(i=1, 2, 3)$ , and the two right-handed neutrinos $\nu_{Rj}\;(j=1, 2)$ as well as the Higgs field $H$ are assigned to certain classes. Then the textures of the lepton mass matrices can be straightforwardly obtained by applying the non-invertible selection rules described in section~\ref{sec:non-invertible-rule}. Since none of the charged lepton masses is vanishing, the rank of the $Y_E$ should be equal to three. We are interested in the lepton mass matrices with vanishing entries, all entries are non-zero or at least one column of the mass matrix are vanishing if three generations of left-handed doublets $\ell_L$ correspond to the same class. Consequently we discard this kind of assignment in the following. Analogously we neglect the cases that either the three right-handed fields $E_{Ri}$ are assigned to the same class.

\subsection{$N=3$}\label{sec:minimalseesaw-m=3}
When $N=3$, there are two classes, $[0]$ and $[1]$. From Eq.~\eqref{eq:fusion-rule-zngagugez2}, we see that the relevant fusion rules are
\begin{eqnarray}
\label{eq:fussionRule-N3}[0]\times[0]=[0] \,,\quad [0]\times[1]=[1] \,,\quad [1]\times[1]=[0]+[1] \,.
\end{eqnarray}
In order to obtain a full-rank charged-lepton mass matrix containing texture zeros, the three generations of left-handed lepton doublets cannot simultaneously correspond to the same conjugacy class. Thus we only have the following possible assignments,
\begin{equation}
\ell_L\sim([0],[0],[1]) \quad\text{or}\quad ([1],[1],[0])
\end{equation}
or their permutations. The above constraint equally applies to the three generations of right-handed charged leptons, which means
\begin{equation}
E_R\sim([0],[0],[1]) \quad\text{or}\quad ([1],[1],[0])
\end{equation}
or their permutations. For simplicity, the two generations of right-handed neutrino are assigned to the following classes
\begin{equation}
\nu_R\sim([0], [1]) \,,
\end{equation}
including their permutations, consequently the right-handed neutrino mass matrix $M_R$ is diagonal, the off-diagonal entry is constrained to be vanishing. The Higgs field can correspond to either $[0]$ or $[1]$. Then, using the fusion rule of Eq.~\eqref{eq:fussionRule-N3}, we can determine the textures of $Y_E$, $M_R$, $Y_{\nu}$ and $M_\nu=-M_DM_R^{-1}M_D^{T}$, and the results are shown in table~\ref{tab:M3-YE-seesaw} and table~\ref{tab:n=3-mrynumnu-seesaw}. It is obvious that the light neutrino mass matrix $M_{\nu}$ is block diagonal or none entry is equal to zero. From table~\ref{tab:M3-YE-seesaw}, we see that four types of full-rank $Y_E$ can be obtained as follow,
\begin{eqnarray}
\label{eq:m=3yetex1}Y_E=\begin{pmatrix}
\times & \times & 0 \\
\times & \times & 0 \\
 0 & 0 & \times
\end{pmatrix} \quad \text{for} \quad (\ell_L;E_R;H)\sim([0],[0],[1];[0],[0],[1];[0]) \,,\nonumber\\
([1],[1],[0];[1],[1],[0];[0]) \,,
\end{eqnarray}
\begin{eqnarray}
\label{eq:m=3yetex2}Y_E=\begin{pmatrix}
\times & \times & 0 \\
\times & \times & 0 \\
\times & \times & \times
\end{pmatrix} \quad \text{for} \quad (\ell_L;E_R;H)\sim([0],[0],[1];[1],[1],[0];[1]) \,,
\end{eqnarray}
\begin{eqnarray}
\label{eq:m=3yetex3}Y_E=\begin{pmatrix}
\times & \times & \times \\
\times & \times & \times \\
0 & 0 & \times
\end{pmatrix} \quad \text{for} \quad (\ell_L;E_R;H)\sim([1],[1],[0];[0],[0],[1];[1]) \,,
\end{eqnarray}
\begin{eqnarray}
\label{eq:m=3yetex4}Y_E=\begin{pmatrix}
\times & \times & \times \\
\times & \times & \times \\
\times & \times & 0
\end{pmatrix} \quad \text{for} \quad (\ell_L;E_R;H)\sim([1],[1],[0];[1],[1],[0];[1]) \,,
\end{eqnarray}
including their permutations of rows and columns. In the simplest scenario of $Z_N$ traditional flavor symmetry, $Q_\psi$ denotes the $Z_N$ charge of the field $\psi$. Obviously the $Z_N$ charges of lepton fields satisfy the following identity,
\begin{equation}
\left(Q_{\bar{\ell}_{Li}}+Q_{E_{Ri}}\right)+\left(Q_{\bar{\ell}_{Lj}}+Q_{E_{Rj}}\right)=\left(Q_{\bar{\ell}_{Li}}+Q_{E_{Rj}}\right)+\left(Q_{\bar{\ell}_{Lj}}+Q_{E_{Ri}}\right)
\end{equation}
for any flavor indices $i, j$. As a consequence, if the off-diagonal $(ij)$ and $(ji)$ entries are non-vanishing, the diagonal entries $(ii)$ and $(jj)$ would be zero or non-zero simultaneously. The same holds true if both $(ij)$ and $(ji)$ entries are vanishing. Thus the sub-blocks $\begin{pmatrix}
\times & \times \\
 \times & 0
\end{pmatrix}$, $\begin{pmatrix}
0 & \times \\
 \times & \times
\end{pmatrix}$, $\begin{pmatrix}
\times & 0 \\
0 & 0
\end{pmatrix}$, $\begin{pmatrix}
0 & 0 \\
0 & \times
\end{pmatrix}$ in the Yukawa matrix is forbidden in the simplest realization of $Z_N$ flavor symmetry\footnote{Notice that these textures could be achievable if ones introduce multiple Higgs fields charged under the Abelian flavor symmetry~\cite{Grimus:2004hf} or there are some flavons in the context of Supersymemtry~\cite{Kageyama:2002zw}.}. It is notable that the $Z_2$ gauging of $Z_N$ symmetry can give rise to certain textures which are inaccessible within the conventional $Z_N$ symmetry. For instance, the textures in Eqs.~(\ref{eq:m=3yetex2}, \ref{eq:m=3yetex3}, \ref{eq:m=3yetex4}) cannot be produced from the $Z_N$ symmetry.

\begin{table}[h]
\centering
\begin{tabular}{|c|c|c|}
	\hline\hline
        & $Y_{E}\,,\,\, H\sim[0]$ & $Y_{E}\,,\,\, H\sim[1]$ \\
	\hline
	\makecell{$\ell_L\sim([0],[0],[1])$ \\ $E_R\sim([0],[0],[1])$} &
	$\begin{pmatrix}
		\times & \times & 0 \\
		\times & \times & 0 \\
		0 & 0 & \times
	\end{pmatrix}$ &
	$\begin{pmatrix}
		0 & 0 & \times\\
		0 & 0 & \times\\
		\times & \times & \times
	\end{pmatrix}$ \\
	\hline
    \makecell{$\ell_L\sim([0],[0],[1])$ \\ $E_R\sim([1],[1],[0])$} &
	$\begin{pmatrix}
		0 & 0 & \times\\
		0 & 0 & \times\\
		\times & \times & 0
	\end{pmatrix}$ &
    $\begin{pmatrix}
		\times & \times & 0 \\
		\times & \times & 0 \\
		\times & \times & \times
	\end{pmatrix}$ \\
	\hline
	\makecell{$\ell_L\sim([1],[1],[0])$ \\ $E_R\sim([0],[0],[1])$} &
	$\begin{pmatrix}
		0 & 0 & \times \\
		0 & 0 & \times \\
		\times & \times & 0
	\end{pmatrix}$ &
	$\begin{pmatrix}
		\times & \times & \times\\
		\times & \times & \times\\
		0 & 0 & \times
	\end{pmatrix}$ \\
	\hline
    \makecell{$\ell_L\sim([1],[1],[0])$ \\ $E_R\sim([1],[1],[0])$} &
	$\begin{pmatrix}
		\times & \times & 0 \\
		\times & \times & 0 \\
		0 & 0 & \times
	\end{pmatrix}$ &
	$\begin{pmatrix}
		\times & \times & \times\\
		\times & \times & \times\\
		\times & \times & 0
	\end{pmatrix}$ \\
	\hline\hline
\end{tabular}
\caption{\label{tab:M3-YE-seesaw}The possible patterns of the charged lepton Yukawa matrix $Y_E$ for different assignments of the fields $\ell_L$, $E_R$ and $H$ in the $Z_2$ gauging of $Z_3$ symmetry, where the symbols ``$\times$'' denotes non-vanishing elements.}
\end{table}

\begin{table}[h]
\centering
\begin{tabular}{|c|c|c|c|c|}
\hline\hline
	& & $M_R$ & $Y_\nu$ & $M_\nu$ \\
\hline
	\multirow{2}{*}[-20pt]{\makecell{$\ell_L\sim([0],[0],[1])$ \\ $\nu_R\sim([0],[1])$}} & $H\sim[0]$ & \multirow{2}{*}[-20pt]{$\begin{pmatrix} \times & 0 \\ 0 & \times \end{pmatrix}$} & $\begin{pmatrix} \times & 0 \\ \times & 0 \\ 0 & \times \end{pmatrix}$ & $\begin{pmatrix} \times & \times & 0 \\ \times & \times & 0 \\ 0 & 0 & \times \end{pmatrix}$ \\
\cline{2-2}\cline{4-5}
    & $H\sim[1]$ & & $\begin{pmatrix} 0 & \times \\ 0 & \times \\ \times & \times \end{pmatrix}$ & $\begin{pmatrix} \times & \times & \times \\ \times & \times & \times \\ \times & \times & \times \end{pmatrix}$ \\
\hline
    \multirow{2}{*}[-20pt]{\makecell{$\ell_L\sim([1],[1],[0])$ \\ $\nu_R\sim([0],[1])$}} & $H\sim[0]$ & \multirow{2}{*}[-20pt]{$\begin{pmatrix} \times & 0 \\ 0 & \times \end{pmatrix}$} & $\begin{pmatrix} 0 & \times \\ 0 & \times \\ \times & 0 \end{pmatrix}$ & $\begin{pmatrix} \times & \times & 0 \\ \times & \times & 0 \\ 0 & 0 & \times \end{pmatrix}$ \\
    \cline{2-2}\cline{4-5}
    & $H\sim[1]$ & & $\begin{pmatrix} \times & \times \\ \times & \times \\ 0 & \times \end{pmatrix}$ & $\begin{pmatrix} \times & \times & \times \\ \times & \times & \times \\ \times & \times & \times \end{pmatrix}$ \\
\hline\hline
\end{tabular}
\caption{\label{tab:n=3-mrynumnu-seesaw}The possible patterns of $Y_{\nu}$, $M_R$ and $M_{\nu}$ for different assignments of $\ell_L$, $\nu_R$ and $H$ in the $Z_2$ gauging of $Z_3$ symmetry, where ``$\times$'' denotes non-zero entry.}
\end{table}

\subsection{$N=4$}\label{sec:minimalseesaw-m=4}
When $N=4$, there are three classes, $[0]$, $[1]$ and $[2]$. The fusion rules are
\begin{eqnarray}
    &&[0]\times[0]=[0] \,,\quad [0]\times[1]=[1] \,,\quad [0]\times[2]=[2] \,,\nonumber\\
    &&[1]\times[1]=[0]+[2] \,,\quad [1]\times[2]=[1] \,,\quad [2]\times[2]=[0] \,.
\end{eqnarray}
The three generations of left-handed lepton doublets can be assigned to the following classes:
\begin{eqnarray}
\ell_L\sim&&([0],[1],[2]) \,,\quad ([0],[0],[1]) \,,\quad ([0],[0],[2]) \,,\quad ([1],[1],[0]) \,,\nonumber\\ &&([1],[1],[2]) \,,\quad ([2],[2],[0]) \,,\quad ([2],[2],[1]) \,,
\end{eqnarray}
up to their permutations. Similarly, the right-handed charged-leptons can correspond to
\begin{eqnarray}
    E_R\sim&&([0],[1],[2]) \,,\quad ([0],[0],[1]) \,,\quad ([0],[0],[2]) \,,\quad ([1],[1],[0]) \,,\nonumber\\ &&([1],[1],[2]) \,,\quad ([2],[2],[0]) \,,\quad ([2],[2],[1]) \,,
\end{eqnarray}
including their permutations. The two generations of right-handed neutrino are assigned to the non-invertible classes in the following patterns:
\begin{equation}
	\nu_R\sim([0], [1]) \,,\quad ([0], [2]) \,,\quad ([1], [2]) \,,
\end{equation}
or their permutations. The Higgs field can correspond to either $[0]$, $[1]$ or $[2]$. The possible patterns of $Y_{\nu}$, $M_R$ and $M_{\nu}$ are summarized in table~\ref{tab:M4-DR-seesaw} and table~\ref{tab:M4-nu-seesaw} respectively. In a similar way, the textures of the charged lepton Yukawa matrix $Y_E$ can be reached\footnote{The possible textures of charged-lepton Yukawa couplings from non-invertible $Z_4$ symmetry have been studied in~\cite{Kobayashi:2025ldi}}.
After enumerating all possible assignments of left- and right-handed lepton fields and Higgs ($7\times7\times3=147$ assignments), we have found two distinct full-rank $Y_E$ textures:
\begin{eqnarray}
    \label{eq:m=4-fullrankye}Y_E=\begin{pmatrix}
        \times & 0 & 0 \\
        0 & \times & 0 \\
        0 & 0 & \times
    \end{pmatrix} \,,\quad \begin{pmatrix}
        \times & \times & 0 \\
        \times & \times & 0 \\
        0 & 0 & \times
    \end{pmatrix} \,,
\end{eqnarray}
up to independent permutations of rows and columns. The corresponding transformations of lepton fields $\ell_L$, $E_R$ and Higgs field $H$ under non-invertible symmetry are given in Appendix~\ref{app:yetex-product}. Note that the above two textures in Eq.~\eqref{eq:m=4-fullrankye} can also be produced by conventional $Z_N$ symmetry. As regard the neutrino sector, the textures of $Y_{\nu}$, $M_R$ and $M_{\nu}$ for different assignments of $\ell_L$, $\nu_R$ and $H$ are collected in Appendix~\ref{app:neutrino-textures-N45}.

\subsection{$N=5$}\label{sec:minimalseesaw-m=5}
When $N=5$, there are three classes $[0]$, $[1]$ and $[2]$. The fusion rules are
\begin{eqnarray}
\label{eq:fussion-rules-N5}&[0]\times[0]=[0] \,,\quad [0]\times[1]=[1] \,,\quad [0]\times[2]=[2] \,,\nonumber\\
&[1]\times[1]=[0]+[2] \,,\quad [1]\times[2]=[1]+[2] \,,\quad [2]\times[2]=[0]+[1] \,.
\end{eqnarray}
It can be observed that the fusion algebra for $N=5$ possesses a symmetry. Specifically, the fusion rules remain invariant under the exchange $[1]\leftrightarrow [2]$. This invariance arises because the group $Z_5$ admits four distinct non-trivial automorphisms, but only a $Z_2$ subgroup of the full automorphism group is gauged.

The three generations of left-handed lepton doublets and right-handed charged-lepton correspond to the classes $[{k_1}]$, $[{k_2}]$, $[{k_3}]$ with $k_1\neq k_2\neq k_3$, there are totally seven possibilities as follows,
\begin{eqnarray}
\ell_L,E_R\sim&&([0],[1],[2]) \,,\quad ([0],[0],[1]) \,,\quad ([0],[0],[2]) \,,\quad ([1],[1],[0]) \,,\nonumber\\ &&([1],[1],[2]) \,,\quad ([2],[2],[0]) \,,\quad ([2],[2],[1]) \,,
\end{eqnarray}
up to permutations. The two generations of right-handed neutrino can be assigned to:
\begin{equation}
\nu_R\sim([0], [1]) \,,\quad ([0], [2]) \,,\quad ([1], [2]) \,,
\end{equation}
including their permutations. The Higgs field $H$ can be $[0]$, $[1]$ or $[2]$. Then, we can examine which entries of $M_R$, $Y_{\nu}$ and $M_\nu=-M_DM_R^{-1}M_D^{T}$ are allowed by our selection rules. The possible textures of $Y_{\nu}$, $M_R$ and $M_{\nu}$ are shown in table~\ref{tab:M5-DR-seesaw} and table~\ref{tab:M5-nu-seesaw} in Appendix~\ref{app:neutrino-textures-N45}. Meanwhile, after enumerating all possible cases of left- and right-handed charged-lepton and higgs ($7\times7\times3=147$ assignments)\footnote{The possible textures of charged-lepton Yukawa couplings non-invertible $Z_5$ symmetry have been studied in~\cite{Kobayashi:2025ldi}}, we have found eight distinct full-rank $Y_E$ textures:
\begin{eqnarray}
\label{eq:m=5-fullrankye}Y_E=&&\begin{pmatrix}
\times & 0 & 0 \\
0 & \times & 0 \\
 0 & 0 & \times
\end{pmatrix} \,,\quad \begin{pmatrix}
0 & \times & 0 \\
\times & 0 & \times \\
0 & \times & \times
\end{pmatrix} \,,\quad \begin{pmatrix}
\times & \times & 0 \\
\times & \times & 0 \\
 0 & 0 & \times
\end{pmatrix} \,,\quad \begin{pmatrix}
0 & 0 & \times \\
\times & \times & 0 \\
\times & \times & \times
\end{pmatrix} \,,\nonumber\\
&&\begin{pmatrix}
0 & \times & \times \\
        0 & \times & \times \\
        \times & 0 & \times
    \end{pmatrix} \,,\quad \begin{pmatrix}
        \times & \times & 0 \\
        \times & \times & 0 \\
        \times & \times & \times
    \end{pmatrix} \,,\quad \begin{pmatrix}
        \times & \times & \times \\
        \times & \times & \times \\
        0 & 0 & \times
    \end{pmatrix} \,,\quad \begin{pmatrix}
        \times & \times & \times \\
        \times & \times & \times \\
        \times & \times & 0
    \end{pmatrix} \,,
\end{eqnarray}
up to permutations of rows and columns. Notice that the second texture in the first line of Eq.~\eqref{eq:m=5-fullrankye} is the so-called
nearest neighbor interaction texture~\cite{Branco:1988iq,Branco:1994jx,Branco:1999nb}. The Appendix~\ref{app:yetex-product} enumerates all possible assignments of the fields $\ell_L$, $E_R$, $H$ which are capable of generating these textures. One sees that the first and third textures in Eq.~\eqref{eq:m=5-fullrankye} can be produced by the traditional $Z_N$ symmetry, while the other textures cannot.

\section{\label{sec:pheno}Textures of lepton mass matrices and phenomenological implications }

In this section, we study the phenomenological implications of texture zeros in $Y_E$, $Y_{\nu}$, and $M_R$ which are derived from the fusion rules of the $Z_2$ gauging of the $Z_N$ symmetry with $N=3,4, 5$. We are interested in the textures of the lepton mass matrices with maximum number of zeros and the second maximum number of zeros, and the charged-lepton masses and neutrino oscillation data are required to be accommodated.

\subsection{Texture zeros of neutrino mass matrix with diagonal charged-lepton mass matrix } \label{sec:diag-charge-lep}

Since none of the three charged leptons are massless, the Yukawa matrix $Y_E$ should be of full-rank. As a result, $Y_E$ can contain at most six texture zeros and it can be taken to be diagonal by using the freedom of field redefinition of lepton fields $\ell_L$ and $E_R$, i.e.
\begin{eqnarray}
\label{eq:YE-diag}Y_E=\begin{pmatrix}
\times & 0 & 0 \\
0 & \times & 0 \\
0 & 0 & \times
\end{pmatrix}\,.
\end{eqnarray}
The textures of $Y_{\nu}$ with three or more zeros for a generic right-handed neutrino mass matrix $M_R$ will lead to vanishing lepton mixing angles or two massless neutrinos, being therefore excluded experimentally. Hence $Y_{\nu}$ can have at most two zero elements. Moreover, textures with two zeros placed in the same row (or column) in $Y_{\nu}$ would lead to two (or one) vanishing lepton mixing angle, they are also excluded by present neutrino data. Thus there are only six $Y_{\nu}$ textures with two zeros~\cite{Barreiros:2018ndn}:
\begin{eqnarray}
\nonumber T_1&=&\begin{pmatrix}
 0 ~&~ \times \\
\times ~&~ 0 \\
\times ~&~ \times
\end{pmatrix} \,,\quad
T_2=\begin{pmatrix}
 0 ~&~ \times \\
 \times ~&~ \times \\
\times ~&~ 0
\end{pmatrix} \,,\quad
T_3=\begin{pmatrix}
\times ~&~ \times \\
0 ~&~ \times \\
\times ~&~ 0
\end{pmatrix} \,,\label{eq:2zeroinYnu-1}\\
T_4&=&\begin{pmatrix}
\times ~&~ 0 \\
 0 ~&~ \times \\
\times ~&~ \times
\end{pmatrix} \,,\quad
T_5=\begin{pmatrix}
\times ~&~ 0 \\
\times ~&~ \times \\
0 ~&~ \times
\end{pmatrix} \,,\quad
T_6=\begin{pmatrix}
\times ~&~ \times \\
\times ~&~ 0 \\
0 ~&~ \times
\end{pmatrix} \,.\label{eq:2zeroinYnu-2}
\end{eqnarray}
From table~\ref{tab:n=3-mrynumnu-seesaw} and table~\ref{tab:M4-DR-seesaw}, we can see that the non-invertible $Z_3$ and $Z_4$ symmetry can not produce any of these six textures $T_1\sim T_6$.

As regards the $Z_2$ gauging of $Z_5$ symmetry, the corresponding fusion rules in Eq.~\eqref{eq:fussion-rules-N5} allow the following three-point couplings, \begin{equation}
\phi_0\phi_0\phi_0 \,,\quad \phi_0\phi_1\phi_1 \,,\quad \phi_0\phi_2\phi_2 \,,\quad \phi_1\phi_1\phi_2 \,,\quad \phi_1\phi_2\phi_2 \,,
\end{equation}
where $\phi_k$ corresponds to the class $[k]$. As shown in table~\ref{tab:M5-DR-seesaw}, all the six textures $T_1\sim T_6$ can be generated from the non-invertible symmetry of $N=5$ for the following assignments of lepton fields and Higgs field,
\begin{eqnarray}
\nonumber T_1&:&\begin{pmatrix}
0 ~&~ \times \\
\times ~&~ 0 \\
\times ~&~ \times
\end{pmatrix} \quad \text{for} \quad (\ell_L;\nu_R;H)\sim([0],[1],[2];[2],[1];[1]) \,,\, ([0],[2],[1];[1],[2];[2]) \,,\label{eq:m=5t1produce}\\
\nonumber T_2&:&
\begin{pmatrix}
0 ~&~ \times \\
\times ~&~ \times \\
\times ~&~ 0
\end{pmatrix} \quad \text{for} \quad (\ell_L;\nu_R;H)\sim([0],[2],[1];[2],[1];[1]) \,,\, ([0],[1],[2];[1],[2];[2]) \,,\label{eq:m=5t2produce}\\
\nonumber  T_3&:&
\begin{pmatrix}
\times ~&~ \times \\
0 ~&~ \times \\
\times ~&~ 0
\end{pmatrix} \quad \text{for} \quad (\ell_L;\nu_R;H)\sim([2],[0],[1];[2],[1];[1]) \,,\, ([1],[0],[2];[1],[2];[2]) \,,\label{eq:m=5t3produce}\\
\nonumber T_4&:&
\begin{pmatrix}
\times ~&~ 0 \\
0 ~&~ \times \\
\times ~&~ \times
\end{pmatrix} \quad \text{for} \quad (\ell_L;\nu_R;H)\sim([0],[1],[2];[1],[2];[1]) \,,\, ([0],[2],[1];[2],[1];[2]) \,,\\
\nonumber T_5&:&
\begin{pmatrix}
\times ~&~ 0 \\
\times ~&~ \times \\
0 ~&~ \times
\end{pmatrix} \quad \text{for} \quad (\ell_L;\nu_R;H)\sim([0],[2],[1];[1],[2];[1]) \,,\, ([0],[1],[2];[2],[1];[2]) \,,\\
 T_6&:&
\begin{pmatrix}
\times ~&~ \times \\
\times ~&~ 0 \\
0 ~&~ \times
\end{pmatrix} \quad \text{for} \quad (\ell_L;\nu_R;H)\sim([2],[0],[1];[1],[2];[1]) \,,\, ([1],[0],[2];[2],[1];[2]) \,. \label{eq:Ynu-T1-T}
\end{eqnarray}
Notice that that the same texture of $Y_{\nu}$ is obtained under the interchange of $[1]$ and $[2]$, since the fusion rules in Eq.~\eqref{eq:fussion-rules-N5} for $N=5$ enjoy this symmetry.

Then we proceed to discuss the texture of the right-handed neutrino mass matrix $M_R$. It is notable that only two textures can be obtained from a generic $Z_2$ gauging of the $Z_N$ symmetries:
\begin{eqnarray}
\nonumber R_1&=&\begin{pmatrix}
\times ~&~ 0 \\
0 ~&~ \times
\end{pmatrix} \quad \text{for} \quad \nu_R\sim([k_1],[k_2])\,, \quad k_1\pm k_2\neq 0 \,,\\
R_2&=&\begin{pmatrix}
\times ~&~ \times \\
\times ~&~ \times
\end{pmatrix} \quad \text{for} \quad \nu_R\sim([k_1],[k_2])\,, \quad k_1\pm k_2=0 \,.\label{eq:R5nurclass}
\end{eqnarray}
From table~\ref{tab:M5-DR-seesaw}, we observe that the texture $R_1$ of the right-handed neutrino mass matrix can really be produced from non-invertible $Z_5$ symmetry for the following assignments of $\nu_R$:
\begin{eqnarray}
R_1=\begin{pmatrix}
\times & 0 \\
 0 & \times
\end{pmatrix} &\text{for}& \nu_R\sim([0],[1]) \,,\, ([0],[2]) \,,\, ([1],[2]) \,,\, ([1],[0]) \,,\, ([2],[0]) \,,\, ([2],[1]) \,.
\end{eqnarray}
However, the second texture $R_2$, when combined with $T_1\sim T_6$, would result in a light neutrino mass matrix $M_\nu=-v^2Y_{\nu}M_R^{-1}Y_{\nu}^T$ which doesn't have vanishing entry. Consequently it is uninteresting to us in light of texture zero.

From Eq.~\eqref{eq:diagYE-N=5}, we see that the desired diagonal charged-lepton Yukawa matrix $Y_E$ can also be obtained from non-invertible $Z_5$ symmetry as:
\begin{equation}
Y_E=\begin{pmatrix}
\times ~&~ 0 ~&~ 0 \\
 0 ~&~ \times ~&~ 0 \\
0 ~&~ 0 ~&~ \times
\end{pmatrix} \quad \text{when} \quad (\ell_L;E_R;H)\sim([0],[1],[2];[0],[1],[2];[0]) \,.
\end{equation}
The Yukawa coupling $Y_E$ is also diagonal if one interchanges the assignments for the three generations of $\ell_L$ and $E_R$ simultaneously. Note that diagonal charged-lepton Yukawa coupling $Y_E$ requires the Higgs to correspond to the class $[0]$, while the neutrino Yukawa coupling $T_1\sim T_6$ demands the Higgs in the class $[1]$ or $[2]$. As a result, one could work in the framework of supersymmetric standard model, in which two Higgs fields $H_u$ and $H_d$ are needed for holomorphicity of supersymmetry and the cancellation of chiral anomalies. $H_u$ and $H_d$ couple with the neutrino and changed lepton Yukawa couplings respectively, and they can correspond to different classes of non-invertible symmetry.

For the texture $R_1$ of $M_R$ and the six textures $T_1\sim T_6$ of $Y_{\nu}$, using the seesaw formula $M_\nu=-\frac{1}{2}v^2Y_{\nu}M_R^{-1}Y_{\nu}^T$ in Eq.~\eqref{eq:eff-neu-mass}, we find that the resulting light neutrino mass matrix has one zero element:
\begin{subequations}
\begin{eqnarray}
\label{eq:Mnu12-zero}&& Y_{\nu}=T_1 ~\text{or}~T_4,~~M_R=R_1,~~M_{\nu}= \begin{pmatrix}
\times ~&~ 0 ~&~ \times \\
0 ~&~ \times ~&~\times \\
\times & \times & \times
\end{pmatrix}\,,\\
\label{eq:Mnu13-zero}&& Y_{\nu}=T_2 ~\text{or}~T_5,~~M_R=R_1,~~M_{\nu}=\begin{pmatrix}
\times ~&~ \times ~&~ 0 \\
\times  ~&~ \times ~&~\times \\
0  ~&~\times ~&~\times
\end{pmatrix}\,,\\
\label{eq:Mnu23-zero}&& Y_{\nu}=T_3 ~\text{or}~T_6,~~M_R=R_1,~~M_{\nu}=\begin{pmatrix}
\times ~&~\times ~&~\times \\
\times ~&~ \times ~&~ 0 \\
\times ~&~ 0 ~&~\times
\end{pmatrix}\,.
\end{eqnarray}
\end{subequations}
One see that the off-diagonal entry $(12)$, $(13)$ or $(23)$ of $M_{\nu}$ is vanishing\footnote{The light neutrino mass matrix $M_{\nu}$ is a symmetric matrix satisfying $(M_{\nu})_{ij}=(M_{\nu})_{ji}$.}. In all these cases, the charged lepton mass matrix is diagonal so that the lepton mixing completely arises from the neutrino sector with $U=U_{\nu_L}$. From Eq.~\eqref{eq:UnuL}, we obtain
\begin{equation}
M_\nu= U\,\text{diag}(m_1, m_2, m_3) U^T\,,
\end{equation}
which implies
\begin{eqnarray}
(M_\nu)_{\alpha\beta}=\sum_{i=1,2,3}m_iU_{\alpha i}U_{\beta i} \,.\label{eq:Mnu-element}
\end{eqnarray}
Consequently the condition of vanishing entry $(M_{\nu})_{\alpha\beta}=0$ imposes relations among the neutrino parameters. To be more specific, we have
\begin{eqnarray}
\text{NO}:&& \quad \frac{m_2}{m_3} = -\frac{U_{\alpha 3} U_{\beta 3}}{U_{\alpha 2} U_{\beta 2}} \,, \nonumber \\
\text{IO}:&& \quad \frac{m_1}{m_2} = -\frac{U_{\alpha 2} U_{\beta 2}}{U_{\alpha 1} U_{\beta 1}} \,, \label{eq:NO-IO-m123}
\end{eqnarray}
in minimal seesaw model, where we have taken into account $m_1=0$ for NO and $m_3=0$ for IO neutrino masses. It is known that the neutrino oscillation experiments can only measure the neutrino mass squared differences $\Delta m^2_{21}\equiv m^2_2-m^2_1$ in solar neutrino experiments and $\Delta m^2_{31}\equiv m^2_3-m^2_1$ in atmospheric neutrino experiments. Thus we have
\begin{eqnarray}
&\text{NO}:& \frac{\Delta m^2_{21}}{\Delta m^2_{31}}=\left| \frac{U_{\alpha 3} U_{\beta 3}}{U_{\alpha 2} U_{\beta 2}} \right|^2 \,, \nonumber \\
&\text{IO}:& \frac{\Delta m^2_{21}}{|\Delta m^2_{31}|}=\left| \frac{U_{\alpha 1} U_{\beta 1}}{U_{\alpha 2} U_{\beta 2}} \right|^2-1 \,. \label{eq:IO-NO-massSQ}
\end{eqnarray}
When the lepton mixing parameters and the mass squared differences $\Delta m^2_{21}$ and $\Delta m^2_{31}$ are limited in the experimentally allowed $3\sigma$ regions~\cite{Esteban_2024}, only the first two cases in Eqs.~(\ref{eq:Mnu12-zero}, \ref{eq:Mnu13-zero}) are viable and the constraint in Eq.~\eqref{eq:IO-NO-massSQ} can be satisfied if the neutrino masses are IO~\cite{Barreiros:2018ndn}. The JUNO experiment is expected to be able to determine the neutrino mass ordering at a $3-4\sigma$ significance after six years running~\cite{JUNO:2015zny}, thus the prediction for IO ordering can be tested in near future. With the standard parametrization of the lepton mixing matrix in Eq.~\eqref{eq:PMNS-para}, for IO neutrino mass spectrum, one can express the Dirac CP violation phase $\cos\delta_{CP}$ in terms of lepton mixing angles and the ratio $\Delta m^2_{21}/|\Delta m^2_{31}|$ as follows
\begin{subequations}
\begin{eqnarray}
\label{eq:Mnu12-cosDelta} (M_{\nu})_{12}=0&:&\cos\delta_{CP}=\frac{[(r_m+1)s^4_{12}-c^4_{12}]s^2_{13}s^2_{23}+r_ms^2_{12}c^2_{12}c^2_{23}}{2(r_ms^2_{12}+1)s_{12}c_{12}s_{13}s_{23}c_{23}} \,,\\
\label{eq:Mnu13-cosDelta} (M_{\nu})_{13}=0&:&\cos\delta_{CP}=-\frac{[(r_m+1)s^4_{12}-c^4_{12}]s^2_{13}c^2_{23}+r_ms^2_{12}c^2_{12}s^2_{23}}{2(r_ms^2_{12}+1)s_{12}c_{12}s_{13}s_{23}c_{23}} \,,\\
\label{eq:Mnu23-cosDelta}(M_{\nu})_{23}=0&:&\cos\delta_{CP}\approx-\frac{[(r_m+1)c_{12}^4-s_{12}^4]c_{12}s_{23}c_{23}}{2s_{12}^3s_{13}\cos2\theta_{23}} \,,
\end{eqnarray}
\end{subequations}
where $r_m\equiv\frac{\Delta m^2_{21}}{|\Delta m^2_{31}|}$. Freely varying $\theta_{12}$, $\theta_{13}$, $\theta_{23}$ and $r_m$ within their $3\sigma$ experimental ranges~\cite{Esteban_2024}, we find that the relations in Eqs.~(\ref{eq:Mnu12-cosDelta},\ref{eq:Mnu13-cosDelta}) constrain the Dirac CP violating phase $\delta_{CP}$ to lie in the following narrow regions\footnote{The relations in Eqs.~(\ref{eq:Mnu12-cosDelta}, \ref{eq:Mnu13-cosDelta}, \ref{eq:Mnu23-cosDelta}) can not fix the sign of $\delta_{CP}$, it is found that $\delta_{CP}$ is limited in the region $\delta_{CP}\in \pm[266.69^{\circ},271.33^{\circ}]$ for $(M_{\nu})_{12}=0$ and $\delta_{CP}\in\pm[268.40^{\circ},272.98^{\circ}]$ for $(M_{\nu})_{13}=0$ respectively. Here the parts of negative sign are dropped in light of the global fitting for $\delta_{CP}$~\cite{Esteban_2024}.}
\begin{subequations}
\begin{eqnarray}
\label{eq:Mnu12-DeltaCP} (M_{\nu})_{12}=0&:&~~266.69^{\circ}\leq\delta_{CP}\leq271.33^{\circ}\,, \\
\label{eq:Mnu13-DeltaCP} (M_{\nu})_{13}=0&:&~~268.40^{\circ}\leq\delta_{CP}\leq272.98^{\circ}\,.
\end{eqnarray}
\end{subequations}
For the texture $(M_{\nu})_{23}=0$,  the relation in Eq.~\eqref{eq:Mnu23-cosDelta} leads to $|\cos\delta_{CP}|>1$ for the experimental values of lepton mixing angles and $r_m$. Hence this texture is not compatible with neutrino data. The predictions for $\delta_{CP}$ in Eqs.~(\ref{eq:Mnu12-DeltaCP}, \ref{eq:Mnu13-DeltaCP}) can be tested by the planned high-precision measurements of neutrino CP violation at the T2HK~\cite{Hyper-Kamiokande:2018ofw} and DUNE~\cite{DUNE:2020ypp} experiments under construction.

\subsection{Texture zeros of neutrino mass matrix with block-diagonal charged-lepton mass matrix}

The preceding sections \ref{sec:minimalseesaw-m=3}, \ref{sec:minimalseesaw-m=4} and \ref{sec:minimalseesaw-m=5} reveal seven distinct types of non-diagonal $Y_E$ textures which can be constructed from non-invertible $Z_{3,4,5}$ symmetries. In the following, we shall discuss the texture
\begin{eqnarray}
\label{eq:YE-blockD}Y_E=\begin{pmatrix}
\times ~& \times ~& 0 \\
\times ~& \times ~& 0 \\
0 ~& 0 ~& \times
\end{pmatrix} \,,\label{eq:blockdiagye}
\end{eqnarray}
and its permutations of rows and columns, which has the next most zeros after the diagonal one.

Since $Y_E$ is block diagonal, the charged lepton sector contributes to one lepton mixing angle. The neutrino Yukawa coupling $Y_{\nu}$ can have at most three zero elements. Textures with three zeros placed in the same column of $Y_{\nu}$ will lead to two massless neutrinos, which is excluded experimentally.
If there are two zeros in some row of $Y_{\nu}$, the light neutrino mass matrix would be block diagonal so that the experimental data of lepton mixing angles can not be accommodated. Therefore there is only one texture of $Y_{\nu}$ with three vanishing entries up to permutations of rows and columns in the minimal seesaw, i.e.
\begin{eqnarray}
 Y^{(3)}_{\nu}=\begin{pmatrix}
\times ~&~ 0 \\
0 ~&~ \times \\
0 ~&~ \times
\end{pmatrix}\,.
\end{eqnarray}
This texture can be obtained from the $Z_2$ gauging of $Z_N$ symmetry for $N=3, 4, 5$, as can be seen from tables~\ref{tab:n=3-mrynumnu-seesaw}, \ref{tab:M4-DR-seesaw} and \ref{tab:M5-DR-seesaw}. We see that the two columns of $Y^{(3)}_{\nu}$ have different textures, the two right-handed neutrinos must carry different charges under the non-invertible symmetry, which forces $M_R$ to be diagonal. As a consequence,
$M_\nu$ is block diagonal as well,
\begin{eqnarray}
\nonumber M_\nu&=&-\frac{1}{2}v^2Y_{\nu}M_R^{-1}Y_{\nu}^T \\
\nonumber &=&\begin{pmatrix}
\times ~&~ 0 \\ 0 ~&~ \times \\ 0 ~&~ \times
\end{pmatrix}\begin{pmatrix}
\times ~&~ 0 \\ 0 ~&~ \times
\end{pmatrix}\begin{pmatrix}
\times ~&~ 0 ~&~ 0 \\ 0 ~&~ \times ~&~ \times
\end{pmatrix} \\
&=&\begin{pmatrix}
\times ~&~ 0 ~&~ 0 \\
0 ~&~ \times ~&~ \times \\
0 ~&~ \times ~&~ \times
\end{pmatrix}\,.
\end{eqnarray}
The block diagonal $Y_E$ and $M_{\nu}$ are excluded by present neutrino data\footnote{If $Y_{E}$ provides a rotation matrix in (12)-plane and $M_\nu$ provides a rotation matrix in (23) plane, the three lepton mixing angles would be correlated as $\tan\theta_{23}\tan\theta_{12}=\sin\theta_{13}$ which is not consistent with the experimental data~\cite{Capozzi:2025wyn}.}.

In the case that $Y_{\nu}$ has two vanishing entries, they can be in the same column or two different columns. Thus $Y_{\nu}$ can be of the following patterns
\begin{eqnarray}
Y^{(2)}_{\nu}=\begin{pmatrix}
\times ~&~ \times \\
0 ~&~ \times \\
0 ~&~ \times
\end{pmatrix} \,,\quad
Y^{'(2)}_{\nu}=\begin{pmatrix}
0 ~&~ \times \\
\times ~&~ 0 \\
\times ~&~ \times
\end{pmatrix}
\end{eqnarray}
up to independent permutations of rows and columns. The texture $Y_E$ in Eq.~\eqref{eq:YE-blockD} and $Y^{(2)}_{\nu}$ can accommodate the measured lepton masses and mixing parameters for a diagonal right-handed neutrino mass matrix $M_R$, and a numerical example in the case of NO is given as follows.
\begin{eqnarray}
\nonumber&& Y_E\frac{v}{\sqrt{2}}=\begin{pmatrix}
44.800+54.000i ~&~ -68.243 ~&~ 0 \\
-14.523-24.146i ~&~ 27.741+4.5401i ~&~ 0 \\
0 ~&~ 0 ~&~ 1776.9
\end{pmatrix}\text{MeV}\,,\\
\label{eq:YE-Ynu-numer}&& Y_{\nu}M^{-1/2}_R\frac{v}{\sqrt{2}}=
\begin{pmatrix}
0.10326+0.039452 i ~&~ -0.10805 \\
0 ~&~ 0.086538 \\
0 ~&~ 0.16738
\end{pmatrix}\sqrt{\text{eV}}\,.
\end{eqnarray}
Following the procedure outlined in section~\ref{sec:texture-zeros-MSM} to diagonalize the corresponding charged lepton mass matrix and light neutrino mass matrix, we obtain the three charged lepton masses as follows,
\begin{equation}
m_e=0.51119\,\text{MeV} \,,\, m_\mu=105.66 \,\text{MeV} \,,\, m_\tau=1776.9 \,\text{MeV} \,,
\end{equation}
which are in well agreement with their measured values~\cite{ParticleDataGroup:2024cfk}, and the light neutrino masses are determined to be
\begin{equation}
m_1=0 \,,\, m_2=8.6544\times10^{-3}\,\text{eV} \,,\, m_3=5.0129\times10^{-2}\text{eV} \,.
\end{equation}
The numerical values of the unitary transformations $U_{\ell_L}$ and  $U_{\nu_L}$ as well as lepton mixing matrix $U=U^{\dagger}_{\ell_L}U_{\nu_L}$ are also be fixed by the diagonalization, and then we can extract the lepton mixing angles and the CP violating phases as,
\begin{equation}
\theta_{12}=33.680\degree \,,\, \theta_{13}=8.5599\degree \,,\, \theta_{23}=43.301\degree \,,\, \delta_{CP}=212.00\degree \,,\,\alpha_{21}=350.54\degree\,.
\end{equation}
All the three mixing angles and Dirac CP phase $\delta_{CP}$ are compatible with the preferred values from global fitting at the $1\sigma$ level~\cite{Esteban_2024}. From the view of non-invertible symmetry, the absence of zeros in the second column of $Y^{(2)}_{\nu}$ requires
\begin{eqnarray}
\label{eq:cantsatiscondi}[k_{\ell_{Li}}]=[k_H+ k_{\nu_{R2}}] ~~\text{or}~~[k_H- k_{\nu_{R2}}]\,,~~i=1,2,3 \,,
\end{eqnarray}
where $[k_{\ell_{Li}}]$, $[k_H]$, $[k_{\nu_{R2}}]$ denote the charges of non-invertible symmetry carried by the left-handed lepton $\ell_{Li}$, Higgs field $H$, and right-handed neutrino $\nu_{R2}$, respectively. As a consequence, the three generations of lepton doublets $\ell_{Li} \;(i=1,2,3)$ have at most two different transformation rules under non-invertible symmetry. Thus only the unviable textures $Y_E=\begin{pmatrix}
\times & \times & 0 \\
\times & \times & 0 \\
 0 & 0 & \times
\end{pmatrix}$, $
Y_{\nu}=\begin{pmatrix}
 0 ~& \times \\
0 ~& \times \\
\times ~& \times
\end{pmatrix}$ and their simultaneous row permutations in this scenario can be produced from the $Z_2$ gauging of $Z_N$ symmetries, yet the phenomenological viable textures such as these in Eq.~\eqref{eq:YE-Ynu-numer} can not, although either $Y_E$ in Eq.~\eqref{eq:YE-blockD} and $Y^{(2)}_{\nu}$ alone can be produced from the non-invertible symmetry.

\begin{figure}[hptb!]
\centering
\includegraphics[scale=0.6]{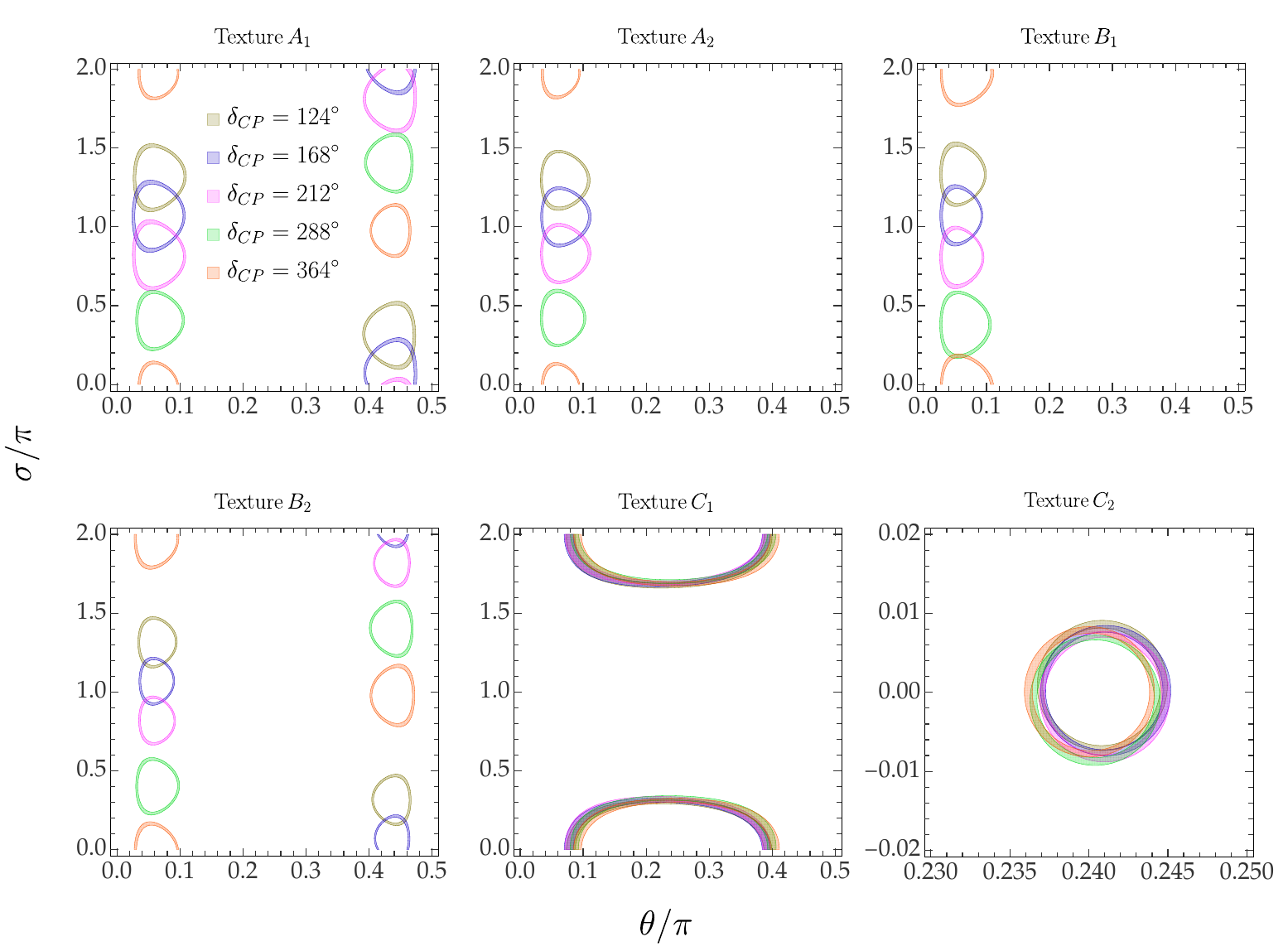}
\caption{The contour plots of the texture zero constraint Eq.~\eqref{eq:NO-io-m221/m231} in the $\theta-\sigma$ plane for the NO neutrino masses. Five representative values $\delta_{CP}=124^{\circ}$, $168^{\circ}$, $212^{\circ}$, $288^{\circ}$, $364^{\circ}$ are shown in different colors, and the width of the lines arises from the $3\sigma$ uncertainty of $\frac{\Delta m^2_{21}}{\Delta m^2_{31}}$~\cite{Esteban_2024}. }
\label{fig:ynud2-no}
\end{figure}

\begin{figure}[hptb!]
\centering
\includegraphics[scale=0.6]{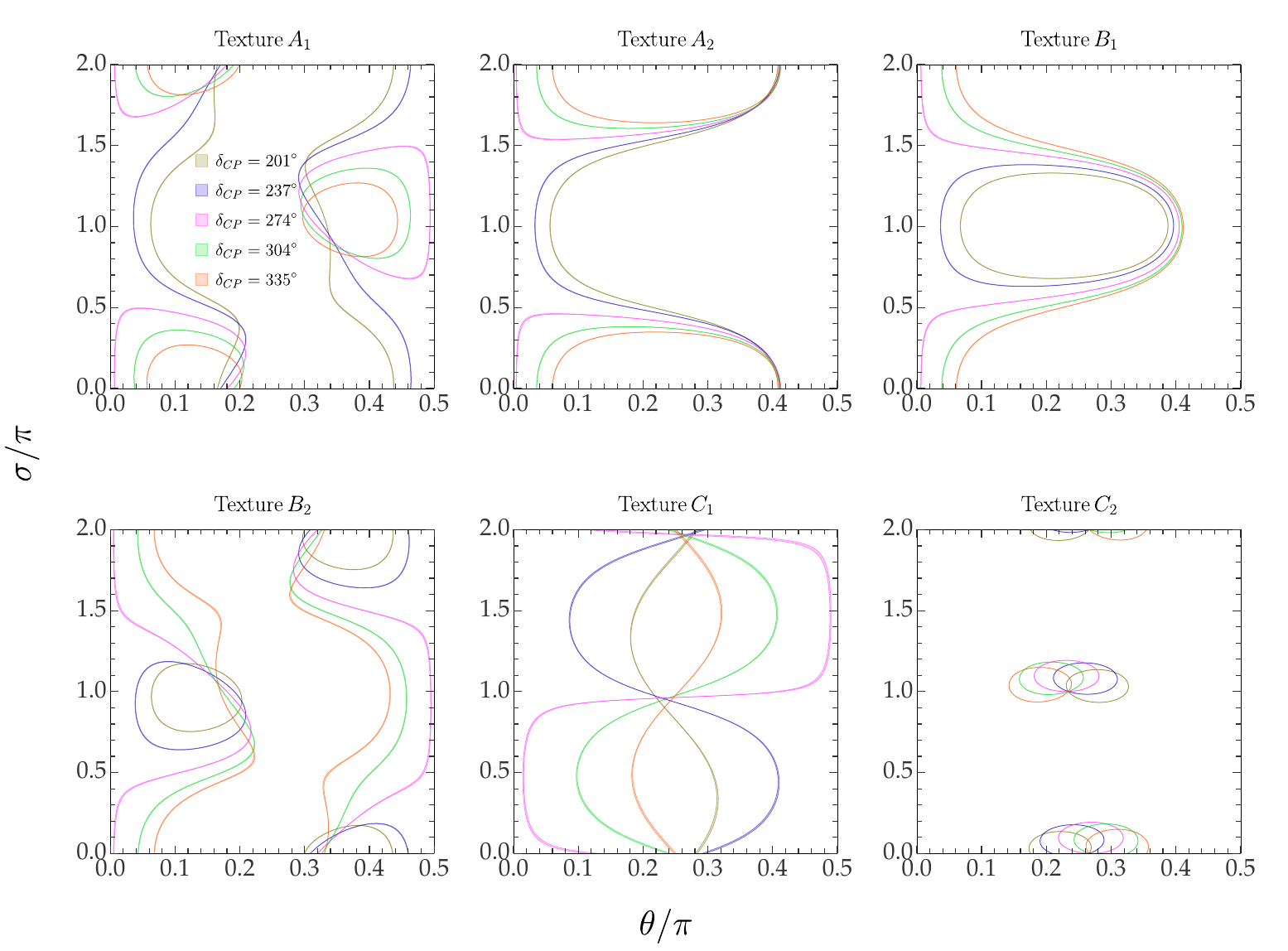}
\caption{The contour plots of the texture zero constraint in Eq.~\eqref{eq:NO-io-m221/m231} in the $\theta-\sigma$ plane for the IO neutrino masses. Five representative values $\delta_{CP}=124^{\circ}$, $168^{\circ}$, $212^{\circ}$, $288^{\circ}$, $364^{\circ}$ are shown in different colors, and the width of the lines arises from the $3\sigma$ uncertainty of $\frac{\Delta m^2_{21}}{|\Delta m^2_{31}|}$~\cite{Esteban_2024}.  }
\label{fig:ynud2-io}
\end{figure}

Now we proceed to consider the remaining alternative structure of the neutrino Yukawa coupling $Y^{'(2)}_{\nu}$. Given the charged lepton Yukawa coupling $Y_E$ in Eq.~\eqref{eq:YE-blockD}, using the freedom of field redefinition, we find that there are six independent textures for the charged lepton and neutrino Yukawa couplings,
\begin{eqnarray}
\nonumber A_1&:& Y_E=\begin{pmatrix} \times ~&~ \times ~&~ 0 \\ \times ~&~ \times ~&~ 0 \\ 0 ~&~ 0 ~&~ \times \end{pmatrix}\,,\quad  Y_{\nu}=\begin{pmatrix} 0 ~&~ \times \\ \times ~&~ 0 \\ \times ~&~ \times \end{pmatrix}\,,\\
\nonumber A_2&:& Y_E=\begin{pmatrix} \times ~&~ \times ~&~ 0 \\ \times ~&~ \times ~&~ 0 \\ 0 ~&~ 0 ~&~ \times \end{pmatrix}\,,\quad  Y_{\nu}=\begin{pmatrix} 0 ~&~ \times \\ \times ~&~ \times \\ \times ~&~ 0 \end{pmatrix}\,,\\
\nonumber B_1&:& Y_E=\begin{pmatrix} \times ~&~ 0 ~&~ \times \\ 0 ~&~ \times ~&~ 0 \\ \times ~&~ 0 ~&~ \times \end{pmatrix}\,,\quad  Y_{\nu}=\begin{pmatrix} 0 ~&~ \times \\ \times ~&~ 0 \\ \times ~&~ \times \end{pmatrix}\,,\\
\nonumber B_2&:& Y_E=\begin{pmatrix} \times ~&~ 0 ~&~ \times \\ 0 ~&~ \times ~&~ 0 \\ \times ~&~ 0 ~&~ \times \end{pmatrix}\,,\quad  Y_{\nu}=\begin{pmatrix} 0 ~&~ \times \\ \times ~&~ \times \\ \times ~&~ 0 \end{pmatrix}\,,\\
\nonumber C_1&:& Y_E=\begin{pmatrix} \times ~&~ 0 ~&~ 0 \\ 0 ~&~ \times ~&~ \times \\ 0 ~&~ \times ~&~ \times \end{pmatrix}\,,\quad  Y_{\nu}=\begin{pmatrix} 0 ~&~ \times \\ \times ~&~ 0 \\ \times ~&~ \times \end{pmatrix}\,,\\
C_2&:& Y_E=\begin{pmatrix} \times ~&~ 0 ~&~ 0 \\ 0 ~&~ \times ~&~ \times \\ 0 ~&~ \times ~&~ \times \end{pmatrix}\,,\quad  Y_{\nu}=\begin{pmatrix} \times ~&~ \times \\ 0 ~&~ \times \\ \times ~&~ 0 \end{pmatrix}\,.\label{eq:notation-ynud2}
\end{eqnarray}
Analogous to the discussions in section~\ref{sec:diag-charge-lep}, the above textures cannot be generated from the non-invertible symmetry for $N=3,4$, while they can be derived from the $Z_2$ gauging of $Z_5$ symmetries. From Eqs.~\eqref{eq:m=5t1produce} and \eqref{eq:m=5yeblockdiag}, we see that the textures $A_1$ and $A_2$ can be naturally obtained if the lepton and Higgs fields transform in the following fashion under the $Z_2$ gauging of $Z_5$ symmetries,
\begin{eqnarray}
A_1: \quad\text{for}\quad(\ell_L;\nu_R;E_R;H_u;H_d)\sim&&([0],[1],[2];[2],[1];[2],[2],[0];[1];[2]) \,,\nonumber\\
&&([0],[2],[1];[1],[2];[1],[1],[0];[2];[1]) \,,\nonumber\\
&&([1],[0],[2];[1],[2];[2],[2],[0];[1];[2]) \,,\nonumber\\
\label{eq:A1-assignment}&&([2],[0],[1];[2],[1];[1],[1],[0];[2];[1]) \,,\\[0.1in]
A_2:\quad\text{for}\quad(\ell_L;\nu_R;E_R;H_u;H_d)\sim&&([0],[2],[1];[2],[1];[1],[1],[0];[1];[1]) \,,\nonumber\\
\label{eq:A2-assignment}&&([0],[1],[2];[1],[2];[2],[2],[0];[2];[2]) \,.
\end{eqnarray}
The charges of the Higgs $H_u$ anb $H_d$ under the non-invertible symmetry are different for $A_1$, while they are the same for $A_2$. Hence the texture $A_1$ can only be produced within the supersymmetric standard model, while $A_2$ no longer necessitates supersymmetry. Similarly the textures $B_1$, $B_2$, $C_1$ and $C_2$ can be derived
by permutating the assignments for three generations of leptons, i.e.
\begin{eqnarray}
B_1: \quad\text{for}\quad(\ell_L;\nu_R;E_R;H_u;H_d)\sim&&([0],[1],[2];[2],[1];[1],[0],[1];[1];[1]) \,,\nonumber\\
\label{eq:B1-assignment}&&([0],[2],[1];[1],[2];[2],[0],[2];[2];[2]) \,,\\[0.1in]
 B_2:\quad\text{for}\quad(\ell_L;\nu_R;E_R;H_u;H_d)\sim&&([0],[2],[1];[2],[1];[2],[0],[2];[1];[2]) \,,\nonumber\\
&&([0],[1],[2];[1],[2];[1],[0],[1];[2];[1]) \,,\nonumber\\
&&([1],[2],[0];[1],[2];[2],[0],[2];[1];[2]) \,,\nonumber\\
\label{eq:B2-assignment}&&([2],[1],[0];[2],[1];[1],[0],[1];[2];[1]) \,,\\ [0.1in]
C_1: \quad\text{for}\quad(\ell_L;\nu_R;E_R;H_u;H_d)\sim&&([1],[0],[2];[1],[2];[0],[1],[1];[1];[1]) \,,\nonumber\\
\label{eq:C1-assignment}&&([2],[0],[1];[2],[1];[0],[2],[2];[2];[2]) \,,\\ [0.1in]
C_2:\quad\text{for}\quad(\ell_L;\nu_R;E_R;H_u;H_d)\sim&&([2],[0],[1];[2],[1];[0],[2],[2];[1];[2]) \,,\nonumber\\
&&([1],[0],[2];[1],[2];[0],[1],[1];[2];[1]) \,,\nonumber\\
&&([2],[1],[0];[1],[2];[0],[2],[2];[1];[2]) \,,\nonumber\\
\label{eq:C2-assignment}&&([1],[2],[0];[2],[1];[0],[1],[1];[2];[1]) \,.
\end{eqnarray}
For all the six cases $A_1$, $A_2$, $B_1$, $B_2$, $C_1$ and $C_2$, the charged lepton mass matrix is block diagonal and it can be diagonalized by a unitary rotation in the (12)-plane, (13)-plane or (23)-plane. Consequently the unitary transformation $U_{\ell L}$ can be parametrized as follows\footnote{Note that we have dropped a diagonal phase matrix in the left side of $U_{\ell L}$ since it is unphysical. },
\begin{eqnarray}
\nonumber A_1~\text{and}~A_2:&&  U_{\ell L}=
\begin{pmatrix}
\cos\theta ~&~ -\sin\theta\,e^{i\sigma} ~&~ 0 \\
\sin\theta\, e^{-i\sigma}~&~ \cos\theta ~&~ 0 \\
0 ~&~ 0 ~&~ 1
\end{pmatrix}\,,\\
\nonumber B_1~\text{and}~B_2:&&  U_{\ell L}=
\begin{pmatrix}
\cos\theta ~&~ 0 ~&~ -\sin\theta\, e^{i\sigma} \\
0  ~&~ 1 ~&~ 0 \\
\sin\theta\, e^{-i\sigma}  ~&~ 0 ~&~ \cos\theta
\end{pmatrix}\,,\\
C_1~\text{and}~C_2:&&  U_{\ell L}=
\begin{pmatrix}
1~&~ 0 ~&~ 0 \\
0~&~ \cos\theta ~&~ -\sin\theta\,e^{i\sigma}  \\
0~&~ \sin\theta\, e^{-i\sigma}~&~ \cos\theta
\end{pmatrix}\,.
\end{eqnarray}
Since the two columns of $Y^{'(2)}_{\nu}$ have different patterns of zeros, two right-handed neutrinos should transform differently under the $Z_2$ gauging of the $Z_N$ symmetries, as can be seen from Eqs.~(\ref{eq:A1-assignment}, \ref{eq:A2-assignment}, \ref{eq:B1-assignment}, \ref{eq:B2-assignment}, \ref{eq:C1-assignment}, \ref{eq:C2-assignment}) as well. Therefore the non-invertible symmetry enforces the right-handed neutrino mass matrix $M_R$ to be diagonal. Using the seesaw formula of Eq.~\eqref{eq:eff-neu-mass}, we find that one element of the light neutrino mass matrix is vanishing. To be more specific, we have $(M_{\nu})_{12}=(M_{\nu})_{21}=0$ for the cases $A_1$, $B_1$, $C_1$, and $(M_{\nu})_{13}=(M_{\nu})_{31}=0$ for the cases $A_2$, $B_2$, and $(M_{\nu})_{23}=(M_{\nu})_{32}=0$ for the case $C_2$. The light neutrino mass matrix $M_{\nu}$ can be expressed in terms of light neutrino masses and lepton mixing matrix as
\begin{eqnarray}
M_\nu=U_{\nu L}\, \text{diag}(m_1, m_2, m_3) U_{\nu L}^T=U_{\ell L} U\, \text{diag}(m_1, m_2, m_3) U^T U_{\ell L}^T\,,
\end{eqnarray}
where the identity $U_{\nu L}=U_{\ell L} U$ has been used. In the minimal seesaw model, one has $m_1=0$ for NO and $m_3=0$ for IO. The vanishing entry $(M_\nu)_{\alpha\beta}=0$ can correlate neutrino masses with mixing parameters as follows
\begin{eqnarray}
\left\{\begin{aligned}
\frac{m_2}{m_3} = -\frac{(U_{\ell L} U)_{\alpha 3} (U_{\ell L} U)_{\beta 3}}{(U_{\ell L} U)_{\alpha 2} (U_{\ell L} U)_{\beta 2}} &\qquad\text{for}\qquad \text{NO} \,,\\
\frac{m_2}{m_1} = -\frac{(U_{\ell L} U)_{\alpha 1} (U_{\ell L} U)_{\beta 1}}{(U_{\ell L} U)_{\alpha 2} (U_{\ell L} U)_{\beta 2}} &\qquad \text{for}\qquad \text{IO} \,.
\end{aligned}
\right.\label{eq:mnu-unul}
\end{eqnarray}
Hence the ratio of the neutrino mass squared differences fulfill the following relations
\begin{eqnarray}
\left\{\begin{aligned}
\frac{\Delta m^2_{21}}{\Delta m^2_{31}} &= \left| \frac{\sum_{i,j}(U_{\ell L})_{\alpha i}U_{i3} (U_{\ell L})_{\beta j}U_{j3}}{\sum_{k,l}(U_{\ell L})_{\alpha k}U_{k2} (U_{\ell L})_{\beta l}U_{l2}} \right|^2 \qquad~~\text{for}\qquad \text{NO} \,,\\
\frac{\Delta m^2_{21}}{|\Delta m^2_{31}|} &= \left|\frac{\sum_{i,j}(U_{\ell L})_{\alpha i}U_{i1} (U_{\ell L})_{\beta j}U_{j1}}{\sum_{k,l}(U_{\ell L})_{\alpha k}U_{k2} (U_{\ell L})_{\beta l}U_{l2}} \right|^2-1\qquad ~~\text{for}\qquad \text{IO} \,.
\end{aligned}
\right.\label{eq:NO-io-m221/m231}
\end{eqnarray}
We list the expressions of $\Delta m^2_{21}/|\Delta m^2_{31}|$ for the six textures $A_i$, $B_i$ and $C_i$ $(i=1, 2)$ in table~\ref{tab:A1-C2NOIOeq}. Due to the presence of $\theta$ and $\sigma$ parameterizing the charged lepton mixing, both NO and IO neutrino masses can be accommodated. We plot the allowed regions of $\theta$ and $\sigma$ constrained by Eq.~\eqref{eq:NO-io-m221/m231} in  figure~\ref{fig:ynud2-no} and figure~\ref{fig:ynud2-io} for NO and IO respectively, where we have taken several representative values of $\delta_{CP}$ in its $3\sigma$ region, and the three mixing angles are fixed at their best-fit values, $\Delta m^2_{21}/|\Delta m^2_{31}|$ is limited within its $3\sigma$ range~\cite{Esteban_2024}.

\begin{table}[t!]
\centering
\begin{tabular}{|c|c|c|}
\hline\hline
$\frac{\Delta m^2_{21}}{|\Delta m^2_{31}|}=$ & NO & IO \\
\hline
$A_1$ & $\left|\frac{(U_{13}^2-U_{23}^2e^{2i\sigma})\sin2\theta+2U_{13}U_{23}e^{i\sigma}\cos2\theta}{(U_{12}^2-U_{22}^2e^{2i\sigma})\sin2\theta+2U_{12}U_{22}e^{i\sigma}\cos2\theta}\right|^2$ & $\left|\frac{(U_{11}^2-U_{21}^2e^{2i\sigma})\sin2\theta+2U_{11}U_{21}e^{i\sigma}\cos2\theta}{(U_{12}^2-U_{22}^2e^{2i\sigma})\sin2\theta+2U_{12}U_{22}e^{i\sigma}\cos2\theta}\right|^2-1$ \\ \hline
$A_2$ & $\left|\frac{U_{33}(U_{13}\cos\theta-U_{23}e^{i\sigma}\sin\theta)}{U_{32}(U_{12}\cos\theta-U_{22}e^{i\sigma}\sin\theta)}\right|^2$ & $\left|\frac{U_{31}(U_{11}\cos\theta-U_{21}e^{i\sigma}\sin\theta)}{U_{32}(U_{12}\cos\theta-U_{22}e^{i\sigma}\sin\theta)}\right|^2-1$ \\ \hline
$B_1$ & $\left|\frac{U_{23}(U_{13}\cos\theta-U_{33}e^{i\sigma}\sin\theta)}{U_{22}(U_{12}\cos\theta-U_{32}e^{i\sigma}\sin\theta)}\right|^2$ & $\left|\frac{U_{21}(U_{11}\cos\theta-U_{31}e^{i\sigma}\sin\theta)}{U_{22}(U_{12}\cos\theta-U_{32}e^{i\sigma}\sin\theta)}\right|^2-1$ \\ \hline
$B_2$ & $\left|\frac{(U_{13}^2-U_{33}^2e^{2i\sigma})\sin2\theta+2U_{13}U_{33}e^{i\sigma}\cos2\theta}{(U_{12}^2-U_{32}^2e^{2i\sigma})\sin2\theta+2U_{12}U_{32}e^{i\sigma}\cos2\theta}\right|^2$ & $\left|\frac{(U_{11}^2-U_{31}^2e^{2i\sigma})\sin2\theta+2U_{11}U_{31}e^{i\sigma}\cos2\theta}{(U_{12}^2-U_{32}^2e^{2i\sigma})\sin2\theta+2U_{12}U_{32}e^{i\sigma}\cos2\theta}\right|^2-1$ \\ \hline
$C_1$ & $\left|\frac{U_{13}(U_{23}\cos\theta-U_{33}e^{i\sigma}\sin\theta)}{U_{12}(U_{22}\cos\theta-U_{32}e^{i\sigma}\sin\theta)}\right|^2$ & $\left|\frac{U_{11}(U_{21}\cos\theta-U_{31}e^{i\sigma}\sin\theta)}{U_{12}(U_{22}\cos\theta-U_{32}e^{i\sigma}\sin\theta)}\right|^2-1$ \\ \hline
$C_2$ & $\left|\frac{(U_{23}^2-U_{33}^2e^{2i\sigma})\sin2\theta+2U_{23}U_{33}e^{i\sigma}\cos2\theta}{(U_{22}^2-U_{32}^2e^{2i\sigma})\sin2\theta+2U_{22}U_{32}e^{i\sigma}\cos2\theta}\right|^2$ & $\left|\frac{(U_{21}^2-U_{31}^2e^{2i\sigma})\sin2\theta+2U_{21}U_{31}e^{i\sigma}\cos2\theta}{(U_{22}^2-U_{32}^2e^{2i\sigma})\sin2\theta+2U_{22}U_{32}e^{i\sigma}\cos2\theta}\right|^2-1$ \\ \hline\hline
\end{tabular}
\caption{\label{tab:A1-C2NOIOeq} The correlation between $\frac{\Delta m^2_{21}}{|\Delta m^2_{31}|}$ and the lepton mixing parameters for the textures $A_1$, $A_2$, $B_1$, $B_2$, $C_1$ and $C_2$, where both NO and IO neutrino mass spectrum are considered. }
\end{table}

%\begin{figure}[t!]

\subsection{The possible new texture zeros from higher $N$}

In this section, we shall investigate whether the $Z_2$ gauging of the $Z_N$ symmetries for larger $N>5$ can give rise to  new textures of $Y_E$ and $M_\nu$. When $N=6, 7$, there are four classes $[0]$, $[1]$, $[2]$ and $[3]$. The three generations of left-handed lepton doublets $\ell_L$ and right-handed charged leptons $E_R$ each admit $=C_6^3-4=16$ distinct charge assignments, where we exclude the assignment that all the three generations $\ell_L$(or $E_R$) correspond to the same class. Similarly the two generations of right-handed neutrino $\nu_R$ admit $C_5^2-4=6$ distinct charge assignments. The Higgs field $H$ can be in $[0]$, $[1]$, $[2]$ or $[3]$.

From Eq.~\eqref{eq:fusion-rule-zngagugez2},  we know that the fusion rules for $N=6$ read as follows,
\begin{eqnarray}
&[0]\times[0]=[0] \,,\quad [0]\times[1]=[1] \,,\quad [0]\times[2]=[2] \,,\quad [0]\times[3]=[3] \,,\nonumber\\
&[1]\times[1]=[0]+[2] \,,\quad [1]\times[2]=[1]+[3] \,,\quad [1]\times[3]=[2] \,,\nonumber\\
&[2]\times[2]=[0]+[2] \,,\quad [2]\times[3]=[1] \,,\quad [3]\times[3]=[0] \,.
\end{eqnarray}
The fusion rules for $N=7$ are different from these of $N=6$ case,
\begin{eqnarray}
&[0]\times[0]=[0] \,,\quad [0]\times[1]=[1] \,,\quad [0]\times[2]=[2] \,,\quad [0]\times[3]=[3] \,,\nonumber\\
&[1]\times[1]=[0]+[2] \,,\quad [1]\times[2]=[1]+[3] \,,\quad [1]\times[3]=[2]+[3] \,,\nonumber\\
&[2]\times[2]=[0]+[3] \,,\quad [2]\times[3]=[1]+[2] \,,\quad [3]\times[3]=[0]+[1] \,.
\end{eqnarray}
We see that the above fusion rules for $N=7$ are invariant under the permutation: $[1]\rightarrow [2]$, $[2]\rightarrow [3]$, $[3]\rightarrow [1]$.

In the case of $N=6$, the corresponding trilinear couplings would be allowed by non-invertible symmetry if one has the following charge assignments for the fields,
\begin{eqnarray}
&[0]\times[0]\times[0] \,,\quad [0]\times[1]\times[1] \,,\quad [0]\times[2]\times[2] \,,\nonumber\\
&[0]\times[3]\times[3] \,,\quad [1]\times[1]\times[2] \,,\quad [1]\times[2]\times[3] \,,\quad [2]\times[2]\times[2] \,.\label{eq:fussion-rules-N6}
\end{eqnarray}
When $N=7$, likewise the charge assignments and contractions for the trilinear couplings compatible with non-invertible symmetry are given by
\begin{eqnarray}
&[0]\times[0]\times[0] \,,\quad [0]\times[1]\times[1] \,,\quad [0]\times[2]\times[2] \,,\quad [0]\times[3]\times[3] \,,\nonumber\\
&[1]\times[1]\times[2] \,,\quad [1]\times[2]\times[3] \,,\quad [1]\times[3]\times[3] \,,\quad [2]\times[2]\times[3] \,. \label{eq:fussion-rules-N7}
\end{eqnarray}
From the selection rules of non-invertible symmetry in Eqs.~(\ref{eq:fussion-rules-N6},\ref{eq:fussion-rules-N7}), we can read out the admissiable trilinear Yukawa couplings. Here we list the textures of full-rank $Y_E$,
\begin{eqnarray}
N=6:~Y_E&=&\begin{pmatrix}
\times & 0 & 0 \\
0 & \times & 0 \\
0 & 0 & \times
\end{pmatrix} \,,~~  \begin{pmatrix}
\times & \times & 0 \\
0 & \times & 0 \\
 0 & 0 & \times
\end{pmatrix} \,,~~  \begin{pmatrix}
\times & \times & 0 \\
\times & \times & 0 \\
 0 & 0 & \times
\end{pmatrix} \,,\nonumber\\
&&\begin{pmatrix}
\times & \times & \times \\
\times & \times & \times \\
\times & 0 & 0
\end{pmatrix} \,,~~  \begin{pmatrix}
\times & \times & \times \\
\times & \times & 0 \\
\times & \times & 0
\end{pmatrix} \,,~~  \begin{pmatrix}
\times & \times & \times \\
\times & \times & \times \\
\times & \times & 0
\end{pmatrix} \,,\label{eq:yetex-n=6} \\
N=7:~Y_E&=&\begin{pmatrix}
\times & 0 & 0 \\
0 & \times & 0 \\
0 & 0 & \times
\end{pmatrix} \,,~~ \begin{pmatrix}
\times & \times & 0 \\
0 & \times & 0 \\
 0 & 0 & \times
\end{pmatrix} \,,~~ \begin{pmatrix}
\times & \times & 0 \\
\times & \times & 0 \\
 0 & 0 & \times
\end{pmatrix} \,,~~ \begin{pmatrix}
\times & \times & 0 \\
\times & 0 & \times \\
 0 & \times & 0
\end{pmatrix} \,,~~\begin{pmatrix}
\times & \times & \times \\
\times & \times & 0 \\
  0 & 0 & \times
\end{pmatrix} \,,\nonumber\\
&& \begin{pmatrix}
\times & \times & 0 \\
\times & \times & 0 \\
\times & 0 & \times
\end{pmatrix} \,,~~  \begin{pmatrix}
\times & \times & \times \\
\times & \times & \times \\
\times & 0 & 0
\end{pmatrix} \,,~~  \begin{pmatrix}
\times & \times & \times \\
\times & \times & 0 \\
\times & \times & 0
\end{pmatrix} \,,~~  \begin{pmatrix}
\times & \times & \times \\
\times & \times & \times \\
\times & \times & 0
\end{pmatrix} \,,\label{eq:yetex-n=7}
\end{eqnarray}
where independent permutations of rows and columns should be included as well. Analogously the following textures of neutrino Yukawa coupling $Y_{\nu}$ of rank two can be achieved,
\begin{eqnarray}
N=6:~Y_\nu&=&\begin{pmatrix}
\times & 0 \\
0 & \times \\
0 & 0
\end{pmatrix} \,,~~ \begin{pmatrix}
\times & 0 \\
\times & 0 \\
0 & \times
\end{pmatrix} \,,~~ \begin{pmatrix}
\times & \times \\
\times & 0 \\
0 & 0
\end{pmatrix} \,,~~\begin{pmatrix}
\times & \times \\
\times & 0 \\
\times & 0
\end{pmatrix} \,,~~ \begin{pmatrix}
\times & \times \\
\times & \times \\
0 & 0
\end{pmatrix} \,,~~ \begin{pmatrix}
\times & \times \\
\times & \times \\
\times & 0
\end{pmatrix} \,,~~~~\quad \label{eq:ynutex-n=6} \\
N=7:~Y_\nu&=&\begin{pmatrix}
\times & 0 \\
 0 & \times \\
 0 & 0
\end{pmatrix} \,,~~\begin{pmatrix}
\times & 0 \\
\times & 0 \\
0 & \times
\end{pmatrix} \,,~~ \begin{pmatrix}
\times & \times \\
\times & 0 \\
0 & 0
\end{pmatrix} \,,~~\begin{pmatrix}
\times & \times \\
\times & 0 \\
\times & 0
\end{pmatrix} \,,\nonumber\\
&&\begin{pmatrix}
\times & \times \\
\times & \times \\
0 & 0
\end{pmatrix} \,,~~ \begin{pmatrix}
\times & \times \\
\times & 0 \\
 0 & \times
\end{pmatrix} \,,~~ \begin{pmatrix}
\times & \times \\
\times & \times \\
\times & 0
\end{pmatrix} \,,\label{eq:ynutex-n=7}
\end{eqnarray}
up to permutations of rows and columns. In Appendix~\ref{app:higher-N-texproduct}, we give an example transformation rule of the fields $\ell_L$, $E_R$, $\nu_R$, $H$ under the $Z_2$ gauging of $Z_N$ symmetries for each texture in above.

In comparison with table~\ref{tab:n=3-mrynumnu-seesaw},  table~\ref{tab:M4-DR-seesaw} and table~\ref{tab:M5-DR-seesaw} for $N=3, 4, 5$ respectively, we see that no new texture of $Y_{\nu}$ is obtained from $N=6, 7$. In fact, there are seven distinct textures of the $3\times2$ matrix $Y_\nu$ with rank 2 and at least one zero entry, up to row and column permutations. All of them can be achieved from the non-invertible $Z_5$ symmetry, as shown in table~\ref{tab:M5-DR-seesaw}. Regarding the charged lepton Yukawa coupling, the $Z_2$ gauging of $Z_N$ symmetries for $N=6, 7$ give rise to a new texture:
\begin{equation}
Y_E=\begin{pmatrix} \times & \times & 0 \\ 0 & \times & 0 \\ 0 & 0 & \times \end{pmatrix}\,.
\end{equation}
When further extending the analysis to $N=9$, one more new texture can be obtained,
\begin{equation}
Y_E=\begin{pmatrix}
\times & \times & 0 \\
\times & 0 & \times \\
0 & \times & \times
\end{pmatrix}\,. \label{eq:yetex-n=9}
\end{equation}
In matrix theory, there exist fourteen distinct textures of full-rank $3\times3$ matrix $Y_E$ containing vanishing elements, up to row and column permutations. Ten of them shown in Eq.~\eqref{eq:yetex-n=7} and Eq.~\eqref{eq:yetex-n=9} can be derived from non-invertible symmetry. The remaining four textures:
\begin{eqnarray}
\begin{pmatrix}
\times & 0 & \times \\
\times & 0 & 0 \\
\times & \times & 0
\end{pmatrix} \,,\quad \begin{pmatrix}
\times & \times & \times \\
0 & 0 & \times \\
\times & 0 & 0
\end{pmatrix} \,,\quad \begin{pmatrix}
\times & \times & \times \\
\times & \times & 0 \\
\times & 0 & 0
\end{pmatrix} \,,\quad \begin{pmatrix}
\times & \times & \times \\
\times & 0 & \times \\
\times & \times & 0
\end{pmatrix}
\end{eqnarray}
cannot be generated from the $Z_2$ gauging of $Z_N$ symmetries, the reason is similar to that discussed in the paragraphs surrounding Eq.~\eqref{eq:cantsatiscondi}.

\section{Conclusion and outlook\label{sec:conclusion}}

The origin of the fermion masses and flavor mixing is a long-standing puzzle in particle physics. Texture zeros refer to vanishing entries in a fermion mass matrix, it is a useful approach to reduce the number of free parameters in mass matrices, thereby enhancing the predictive power of a model. The zero entries in fermion mass matrices could lead to testable relations between fermion mass ratios and mixing parameters. The texture zeros of neutrino mass matrix have been extensively studied to explain observed neutrino oscillation data and to predict yet-undetermined quantities such as the leptonic CP violation phase, the neutrino mass ordering, and the absolute neutrino mass scale~\cite{Frampton:2002yf,Xing:2002ta,Fritzsch:1999ee,Gupta:2012fsl}.

From the theoretical viewpoint, the texture zeros may arise from some continuous or discrete flavor symmetries~\cite{Feruglio:2019ybq,Xing:2020ijf,Ding:2023htn,Ding:2024ozt}, they may
reflect some underlying selection rules. Recently it was found that the $Z_2$ gauging of discrete
$Z_N$ symmetries can constrains the structure of the quark and lepton mass matrices, leading to specific patterns of vanishing entries~\cite{Kobayashi:2024cvp,Kobayashi:2025znw,Kobayashi:2025ldi,Liang:2025dkm,Kobayashi:2025thd}. In this paper, we have performed a systematical study of the texture zeros of the lepton mass matrices in the context of minimal seesaw model in which only two right-handed neutrinos are introduced to generate tiny neutrino masses. We study the textures of the charged lepton Yukawa coupling $Y_E$, the neutrino Yukawa
coupling $Y_{\nu}$, the right-handed neutrino mass matrix $M_R$ and the light neutrino mass matrix $M_{\nu}$ which can be derived from the $Z_2$ gauging of $Z_N$ symmetry. We are particularly interested in the phenomenologically viable textures with the maximum number of zeros and the second maximum number of zeros.

In the charged lepton diagonal basis, the neutrino Yukawa coupling $Y_{\nu}$ can have at most two zero elements to be compatible with experimental data and there are totally six distinct textures $T_{1, 2, 3, 4, 5, 6}$, as shown in Eq.~\eqref{eq:Ynu-T1-T}. For proper assignments of matter fields, the $Z_2$ gauging of $Z_5$ symmetry can give rise to all the six textures $T_{1, 2, 3, 4, 5, 6}$ together with a diagonal $Y_E$, and the right-handed neutrino mass matrix $M_R$ is enforced to be diagonal. As a result, one entry of the light neutrino mass matrix $M_{\nu}$ is vanishing. This zero entry allows to relate the Dirac CP phase $\delta_{CP}$ with the lepton mixing angles and the ratio $r_m\equiv\Delta m^2_{21}/|\Delta m^2_{31}|$. It turns out that only the textures $T_1$, $T_2$, $T_4$ and $T_5$ are compatible with the current neutrino oscillation data, and $\delta_{CP}$ is constrained in narrow regions shown in Eqs.~(\ref{eq:Mnu12-DeltaCP}, \ref{eq:Mnu13-DeltaCP}).

Besides the diagonal charged lepton Yukawa coupling, the $Z_2$ gauging of $Z_N$ symmetry can also generate block diagonal $Y_E$. Taking into account the neutrino Yukawa coupling $Y_{\nu}$, six textures of $Y_E$ and $Y_\nu$ named as $A_{1,2}$, $B_{1,2}$ and $C_{1,2}$ can be obtained from non-invertible $Z_5$ symmetry, as shown in Eqs.~(\ref{eq:A1-assignment},\ref{eq:A2-assignment}, \ref{eq:B1-assignment}, \ref{eq:B2-assignment}, \ref{eq:C1-assignment}, \ref{eq:C2-assignment}). In this scenario, the charged lepton sector contributes a unitary rotation in the $(ij)$-plane ($i, j=1, 2, 3$ and $i<j$) to the lepton mixing, consequently both NO and IO neutrino mass spectrums can be accommodated. Furthermore, the analysis is extended to the large $N$ case. It is found that two more new textures $Y_E=\begin{pmatrix} \times & \times & 0 \\ 0 & \times & 0 \\ 0 & 0 & \times \end{pmatrix}$ and $Y_E=\begin{pmatrix}
\times & \times & 0 \\
\times & 0 & \times \\
0 & \times & \times
\end{pmatrix}$ can be generated from the non-invertible $Z_{6,7}$ and $Z_9$ symmetries respectively. Notice that the former pattern of $Y_E$ has five zero entries. However, no new texture of $Y_{\nu}$ can be produced anymore.

It is intriguing that the $Z_2$ gauging of $Z_N$ symmetry can naturally constrain certain elements of the quark and lepton mass matrices to vanish, resulting in predictive textures which can be tested against the more precise experimental data in near future. In particularly, it can generate some textures which can not be obtained from the traditional $Z_N$ or $U(1)$ abelian flavor symmetry. It is known that the quarks and leptons fields are embedded into few gauge multiplets in Grand Unification Theory so that the quark and lepton mass matrices are related to each other. It is interesting to study what kinds of textures of fermion mass matrices can be generated from the non-invertible $Z_N$ symmetry in the context of Grand Unification Theory, we expect to get more predictive textures which are constrained by the abundant experimental data of both neutrino oscillation and rare meson decays. It is also quite interesting to explore the phenomenological implications of the texture zeros derived from non-invertible symmetry in neutrinoless double decay, strong CP problem and baryon asymmetry of the Universe and so on. All these are left for future.

\section*{Acknowledgements}

This work is supported by the National Natural Science Foundation of China under Grant No.~12375104.

\clearpage

\begin{appendix}

\section{\label{app:neutrino-textures-N45}Possible textures of $Y_{\nu}$, $M_R$ and $M_{\nu}$ for $N=4, 5$}

\subsection{ $N=4$}

Based on the $Z_2$ gauging of $Z_4$ symmetry in minimal seesaw model, we report the textures of the neutrino Yukawa matrix $Y_{\nu}$ and the right-handed neutrino mass matrix $M_{R}$ in table~\ref{tab:M4-DR-seesaw}. The different cases are classified according to the transformation of $\ell_L$, $\nu_R$ and $H$ under the non-invertible symmetry. The corresponding patterns of the light neutrino mass matrix $M_{\nu}$ are listed in table~\ref{tab:M4-nu-seesaw}. Notice that  interchanging the assignment of $\ell_L$ leads to row permutation of $Y_{\nu}$ and $M_{\nu}$, while interchanging the assignment of $\nu_R$ leads to column permutation of $Y_{\nu}$ and $M_{R}$.

\begin{longtable}{|c|c|c|c|c|}
    \midrule\hline
    & $Y_{\nu}\,,\,\, H\sim[0]$ & $Y_{\nu}\,,\,\, H\sim[1]$ & $Y_{\nu}\,,\,\, H\sim[2]$ & $M_R$ \\
    \hline
    \endfirsthead
	
    \hline
    & $Y_{\nu}\,,\,\, H\sim[0]$ & $Y_{\nu}\,,\,\, H\sim[1]$ & $Y_{\nu}\,,\,\, H\sim[2]$ & $M_R$ \\
    \hline
    \endhead
	
        \makecell{$\ell_L\sim([0],[1],[2])$ \\ $\nu_R\sim([0],[1])$} &
        $\begin{pmatrix}
	   \times & 0 \\
	   0 & \times \\
	   0 & 0
        \end{pmatrix}$ &
	$\begin{pmatrix}
		0 & \times\\
		\times & 0\\
		0 & \times
	\end{pmatrix}$ &
	$\begin{pmatrix}
		0 & 0 \\
		0 & \times\\
		\times & 0
	\end{pmatrix}$ &
	$\begin{pmatrix}
		\times & 0 \\
		0 & \times
	\end{pmatrix}$ \\
	\hline
	\makecell{$\ell_L\sim([0],[1],[2])$ \\ $\nu_R\sim([0],[2])$} &
	$\begin{pmatrix}
		\times & 0 \\
		0 & 0 \\
		0 & \times
	\end{pmatrix}$ &
	$\begin{pmatrix}
		0 & 0\\
		\times & \times\\
		0 & 0
	\end{pmatrix}$ &
	$\begin{pmatrix}
		0 & \times \\
		0 & 0\\
		\times & 0
	\end{pmatrix}$ &
	$\begin{pmatrix}
		\times & 0 \\
		0 & \times
	\end{pmatrix}$ \\
	\hline
	\makecell{$\ell_L\sim([0],[1],[2])$ \\ $\nu_R\sim([1],[2])$} &
	$\begin{pmatrix}
		0 & 0 \\
		\times & 0 \\
		0 & \times
	\end{pmatrix}$ &
	$\begin{pmatrix}
		\times & 0\\
		0 & \times\\
		\times & 0
	\end{pmatrix}$ &
	$\begin{pmatrix}
		0 & \times \\
		\times & 0\\
		0 & 0
	\end{pmatrix}$ &
	$\begin{pmatrix}
		\times & 0 \\
		0 & \times
	\end{pmatrix}$ \\
	\hline
	\makecell{$\ell_L\sim([0],[0],[1])$ \\ $\nu_R\sim([0],[1])$} &
	$\begin{pmatrix}
		\times & 0 \\
		\times & 0 \\
		0 & \times
	\end{pmatrix}$ &
	$\begin{pmatrix}
		0 & \times\\
		0 & \times\\
		\times & 0
	\end{pmatrix}$ &
	$\begin{pmatrix}
		0 & 0 \\
		0 & 0\\
		0 & \times
	\end{pmatrix}$ &
	$\begin{pmatrix}
		\times & 0 \\
		0 & \times
	\end{pmatrix}$ \\
	\hline
	\makecell{$\ell_L\sim([0],[0],[1])$ \\ $\nu_R\sim([0],[2])$} &
	$\begin{pmatrix}
		\times & 0 \\
		\times & 0 \\
		0 & 0
	\end{pmatrix}$ &
	$\begin{pmatrix}
		0 & 0\\
		0 & 0\\
		\times & \times
	\end{pmatrix}$ &
	$\begin{pmatrix}
		0 & \times \\
		0 & \times\\
		0 & 0
	\end{pmatrix}$ &
	$\begin{pmatrix}
		\times & 0 \\
		0 & \times
	\end{pmatrix}$ \\
	\hline
	\makecell{$\ell_L\sim([0],[0],[1])$ \\ $\nu_R\sim([1],[2])$} &
	$\begin{pmatrix}
		0 & 0 \\
		0 & 0 \\
		\times & 0
	\end{pmatrix}$ &
	$\begin{pmatrix}
		\times & 0 \\
		\times & 0 \\
		0 & \times
	\end{pmatrix}$ &
	$\begin{pmatrix}
		0 & \times \\
		0 & \times \\
		\times & 0
	\end{pmatrix}$ &
	$\begin{pmatrix}
		\times & 0 \\
		0 & \times
	\end{pmatrix}$ \\
	\hline
	\makecell{$\ell_L\sim([0],[0],[2])$ \\ $\nu_R\sim([0],[1])$} &
	$\begin{pmatrix}
		\times & 0 \\
		\times & 0 \\
		0 & 0
	\end{pmatrix}$ &
	$\begin{pmatrix}
		0 & \times\\
		0 & \times\\
		0 & \times
	\end{pmatrix}$ &
	$\begin{pmatrix}
		0 & 0 \\
		0 & 0\\
		\times & 0
	\end{pmatrix}$ &
	$\begin{pmatrix}
		\times & 0 \\
		0 & \times
	\end{pmatrix}$ \\
	\hline
	\makecell{$\ell_L\sim([0],[0],[2])$ \\ $\nu_R\sim([0],[2])$} &
	$\begin{pmatrix}
		\times & 0 \\
		\times & 0 \\
		0 & \times
	\end{pmatrix}$ &
	$\begin{pmatrix}
		0 & 0 \\
		0 & 0 \\
		0 & 0
	\end{pmatrix}$ &
	$\begin{pmatrix}
		0 & \times \\
		0 & \times \\
		\times & 0
	\end{pmatrix}$ &
	$\begin{pmatrix}
		\times & 0 \\
		0 & \times
	\end{pmatrix}$ \\
	\hline
	\makecell{$\ell_L\sim([0],[0],[2])$ \\ $\nu_R\sim([1],[2])$} &
	$\begin{pmatrix}
		0 & 0 \\
		0 & 0 \\
		0 & \times
	\end{pmatrix}$ &
	$\begin{pmatrix}
		\times & 0 \\
		\times & 0 \\
		\times & 0
	\end{pmatrix}$ &
	$\begin{pmatrix}
		0 & \times \\
		0 & \times \\
		0 & 0
	\end{pmatrix}$ &
	$\begin{pmatrix}
		\times & 0 \\
		0 & \times
	\end{pmatrix}$ \\
	\hline
	\makecell{$\ell_L\sim([1],[1],[0])$ \\ $\nu_R\sim([0],[1])$} &
	$\begin{pmatrix}
		0 & \times \\
		0 & \times \\
		\times & 0
	\end{pmatrix}$ &
	$\begin{pmatrix}
		\times & 0 \\
		\times & 0 \\
		0 & \times
	\end{pmatrix}$ &
	$\begin{pmatrix}
		0 & \times \\
		0 & \times \\
		0 & 0
	\end{pmatrix}$ &
	$\begin{pmatrix}
		\times & 0 \\
		0 & \times
	\end{pmatrix}$ \\
	\hline
	\makecell{$\ell_L\sim([1],[1],[0])$ \\ $\nu_R\sim([0],[2])$} &
	$\begin{pmatrix}
		0 & 0 \\
		0 & 0 \\
		\times & 0
	\end{pmatrix}$ &
	$\begin{pmatrix}
		\times & \times \\
		\times & \times \\
		0 & 0
	\end{pmatrix}$ &
	$\begin{pmatrix}
		0 & 0 \\
		0 & 0 \\
		0 & \times
	\end{pmatrix}$ &
	$\begin{pmatrix}
		\times & 0 \\
		0 & \times
	\end{pmatrix}$ \\
	\hline
	\makecell{$\ell_L\sim([1],[1],[0])$ \\ $\nu_R\sim([1],[2])$} &
	$\begin{pmatrix}
		\times & 0 \\
		\times & 0 \\
		0 & 0
	\end{pmatrix}$ &
	$\begin{pmatrix}
		0 & \times \\
		0 & \times \\
		\times & 0
	\end{pmatrix}$ &
	$\begin{pmatrix}
		\times & 0 \\
		\times & 0 \\
		0 & \times
	\end{pmatrix}$ &
	$\begin{pmatrix}
		\times & 0 \\
		0 & \times
	\end{pmatrix}$ \\
	\hline
	\makecell{$\ell_L\sim([1],[1],[2])$ \\ $\nu_R\sim([0],[1])$} &
	$\begin{pmatrix}
		0 & \times \\
		0 & \times \\
		0 & 0
	\end{pmatrix}$ &
	$\begin{pmatrix}
		\times & 0 \\
		\times & 0 \\
		0 & \times
	\end{pmatrix}$ &
	$\begin{pmatrix}
		0 & \times \\
		0 & \times \\
		\times & 0
	\end{pmatrix}$ &
	$\begin{pmatrix}
		\times & 0 \\
		0 & \times
	\end{pmatrix}$ \\
	\hline
	\makecell{$\ell_L\sim([1],[1],[2])$ \\ $\nu_R\sim([0],[2])$} &
	$\begin{pmatrix}
		0 & 0 \\
		0 & 0 \\
		0 & \times
	\end{pmatrix}$ &
	$\begin{pmatrix}
		\times & \times \\
		\times & \times \\
		0 & 0
	\end{pmatrix}$ &
	$\begin{pmatrix}
		0 & 0 \\
		0 & 0 \\
		\times & 0
	\end{pmatrix}$ &
	$\begin{pmatrix}
		\times & 0 \\
		0 & \times
	\end{pmatrix}$ \\
	\hline
	\makecell{$\ell_L\sim([1],[1],[2])$ \\ $\nu_R\sim([1],[2])$} &
	$\begin{pmatrix}
		\times & 0 \\
		\times & 0 \\
		0 & \times
	\end{pmatrix}$ &
	$\begin{pmatrix}
		0 & \times \\
		0 & \times \\
		\times & 0
	\end{pmatrix}$ &
	$\begin{pmatrix}
		\times & 0 \\
		\times & 0 \\
		0 & 0
	\end{pmatrix}$ &
	$\begin{pmatrix}
		\times & 0 \\
		0 & \times
	\end{pmatrix}$ \\
	\hline
	\makecell{$\ell_L\sim([2],[2],[0])$ \\ $\nu_R\sim([0],[1])$} &
	$\begin{pmatrix}
		0 & 0 \\
		0 & 0 \\
		\times & 0
	\end{pmatrix}$ &
	$\begin{pmatrix}
		0 & \times \\
		0 & \times \\
		0 & \times
	\end{pmatrix}$ &
	$\begin{pmatrix}
		\times & 0 \\
		\times & 0 \\
		0 & 0
	\end{pmatrix}$ &
	$\begin{pmatrix}
		\times & 0 \\
		0 & \times
	\end{pmatrix}$ \\
	\hline
	\makecell{$\ell_L\sim([2],[2],[0])$ \\ $\nu_R\sim([0],[2])$} &
	$\begin{pmatrix}
		0 & \times \\
		0 & \times \\
		\times & 0
	\end{pmatrix}$ &
	$\begin{pmatrix}
		0 & 0 \\
		0 & 0 \\
		0 & 0
	\end{pmatrix}$ &
	$\begin{pmatrix}
		\times & 0 \\
		\times & 0 \\
		0 & \times
	\end{pmatrix}$ &
	$\begin{pmatrix}
		\times & 0 \\
		0 & \times
	\end{pmatrix}$ \\
	\hline
	\makecell{$\ell_L\sim([2],[2],[0])$ \\ $\nu_R\sim([1],[2])$} &
	$\begin{pmatrix}
		0 & \times \\
		0 & \times \\
		0 & 0
	\end{pmatrix}$ &
	$\begin{pmatrix}
		\times & 0 \\
		\times & 0 \\
		\times & 0
	\end{pmatrix}$ &
	$\begin{pmatrix}
		0 & 0 \\
		0 & 0 \\
		0 & \times
	\end{pmatrix}$ &
	$\begin{pmatrix}
		\times & 0 \\
		0 & \times
	\end{pmatrix}$ \\
	\hline
	\makecell{$\ell_L\sim([2],[2],[1])$ \\ $\nu_R\sim([0],[1])$} &
	$\begin{pmatrix}
		0 & 0 \\
		0 & 0 \\
		0 & \times
	\end{pmatrix}$ &
	$\begin{pmatrix}
		0 & \times \\
		0 & \times \\
		\times & 0
	\end{pmatrix}$ &
	$\begin{pmatrix}
		\times & 0 \\
		\times & 0 \\
		0 & \times
	\end{pmatrix}$ &
	$\begin{pmatrix}
		\times & 0 \\
		0 & \times
	\end{pmatrix}$ \\
	\hline
	\makecell{$\ell_L\sim([2],[2],[1])$ \\ $\nu_R\sim([0],[2])$} &
	$\begin{pmatrix}
		0 & \times \\
		0 & \times \\
		0 & 0
	\end{pmatrix}$ &
	$\begin{pmatrix}
		0 & 0 \\
		0 & 0 \\
		\times & \times
	\end{pmatrix}$ &
	$\begin{pmatrix}
		\times & 0 \\
		\times & 0 \\
		0 & 0
	\end{pmatrix}$ &
	$\begin{pmatrix}
		\times & 0 \\
		0 & \times
	\end{pmatrix}$ \\
	\hline
	\makecell{$\ell_L\sim([2],[2],[1])$ \\ $\nu_R\sim([1],[2])$} &
	$\begin{pmatrix}
		0 & \times \\
		0 & \times \\
		\times & 0
	\end{pmatrix}$ &
	$\begin{pmatrix}
		\times & 0 \\
		\times & 0 \\
		0 & \times
	\end{pmatrix}$ &
	$\begin{pmatrix}
		0 & 0 \\
		0 & 0 \\
		\times & 0
	\end{pmatrix}$ &
	$\begin{pmatrix}
		\times & 0 \\
		0 & \times
	\end{pmatrix}$ \\
	\hline\midrule
\caption{\label{tab:M4-DR-seesaw} $Y_{\nu}$ and $M_R$ matrices in minimal seesaw from $Z_2$ gauging of $Z_4$ symmetries, where the crosses represent non-zero matrix elements. }
\end{longtable}

\begin{longtable}{|c|c|c|c|}
	\midrule\hline
	& $M_\nu\,,\,\, H\sim[0]$ & $M_\nu\,,\,\, H\sim[1]$ & $M_\nu\,,\,\, H\sim[2]$ \\
	\hline
	\endfirsthead
		
	\hline
	& $M_\nu\,,\,\, H\sim[0]$ & $M_\nu\,,\,\, H\sim[1]$ & $M_\nu\,,\,\, H\sim[2]$ \\
	\hline
	\endhead
		
	\makecell{$\ell_L\sim([0],[1],[2])$ \\ $\nu_R\sim([0],[1])$} &
	$\begin{pmatrix}
		\times & 0 & 0\\
		0 & \times & 0\\
		0 & 0 & 0
	\end{pmatrix}$ &
	$\begin{pmatrix}
		\times & 0 & \times\\
		0 & \times & 0\\
		\times & 0 & \times
	\end{pmatrix}$ &
	$\begin{pmatrix}
		0 & 0 & 0\\
		0 & \times & 0\\
		0 & 0 & \times
	\end{pmatrix}$ \\
	\hline
	\makecell{$\ell_L\sim([0],[1],[2])$ \\ $\nu_R\sim([0],[2])$} &
	$\begin{pmatrix}
		\times & 0 & 0\\
		0 & 0 & 0\\
		0 & 0 & \times
	\end{pmatrix}$ &
	$\begin{pmatrix}
		0 & 0 & 0\\
		0 & \times & 0\\
		0 & 0 & 0
	\end{pmatrix}$ &
	$\begin{pmatrix}
		\times & 0 & 0\\
		0 & 0 & 0\\
		0 & 0 & \times
	\end{pmatrix}$ \\
	\hline
	\makecell{$\ell_L\sim([0],[1],[2])$ \\ $\nu_R\sim([1],[2])$} &
	$\begin{pmatrix}
		0 & 0 & 0\\
		0 & \times & 0\\
		0 & 0 & \times
	\end{pmatrix}$ &
	$\begin{pmatrix}
		\times & 0 & \times\\
		0 & \times & 0\\
		\times & 0 & \times
	\end{pmatrix}$ &
	$\begin{pmatrix}
		\times & 0 & 0\\
		0 & \times & 0\\
		0 & 0 & 0
	\end{pmatrix}$ \\
	\hline
	\makecell{$\ell_L\sim([0],[0],[1])$ \\ $\nu_R\sim([0],[1])$} &
	$\begin{pmatrix}
		\times & \times & 0\\
		\times & \times & 0\\
		0 & 0 & \times
	\end{pmatrix}$ &
	$\begin{pmatrix}
		\times & \times & 0\\
		\times & \times & 0\\
		0 & 0 & \times
	\end{pmatrix}$ &
	$\begin{pmatrix}
		0 & 0 & 0\\
		0 & 0 & 0\\
		0 & 0 & \times
	\end{pmatrix}$ \\
	\hline
	\makecell{$\ell_L\sim([0],[0],[1])$ \\ $\nu_R\sim([0],[2])$} &
	$\begin{pmatrix}
		\times & \times & 0\\
		\times & \times & 0\\
		0 & 0 & 0
	\end{pmatrix}$ &
	$\begin{pmatrix}
		0 & 0 & 0\\
		0 & 0 & 0\\
		0 & 0 & \times
	\end{pmatrix}$ &
	$\begin{pmatrix}
		\times & \times & 0\\
		\times & \times & 0\\
		0 & 0 & 0
	\end{pmatrix}$ \\
	\hline
	\makecell{$\ell_L\sim([0],[0],[1])$ \\ $\nu_R\sim([1],[2])$} &
	$\begin{pmatrix}
		0 & 0 & 0\\
		0 & 0 & 0\\
		0 & 0 & \times
	\end{pmatrix}$ &
	$\begin{pmatrix}
		\times & \times & 0\\
		\times & \times & 0\\
		0 & 0 & \times
	\end{pmatrix}$ &
	$\begin{pmatrix}
		\times & \times & 0\\
		\times & \times & 0\\
		0 & 0 & \times
	\end{pmatrix}$ \\
	\hline
	\makecell{$\ell_L\sim([0],[0],[2])$ \\ $\nu_R\sim([0],[1])$} &
	$\begin{pmatrix}
		\times & \times & 0\\
		\times & \times & 0\\
		0 & 0 & 0
	\end{pmatrix}$ &
	$\begin{pmatrix}
		\times & \times & \times\\
		\times & \times & \times\\
		\times & \times & \times
	\end{pmatrix}$ &
	$\begin{pmatrix}
		0 & 0 & 0\\
		0 & 0 & 0\\
		0 & 0 & \times
	\end{pmatrix}$ \\
	\hline
	\makecell{$\ell_L\sim([0],[0],[2])$ \\ $\nu_R\sim([0],[2])$} &
	$\begin{pmatrix}
		\times & \times & 0\\
		\times & \times & 0\\
		0 & 0 & \times
	\end{pmatrix}$ &
	$\begin{pmatrix}
		0 & 0 & 0\\
		0 & 0 & 0\\
		0 & 0 & 0
	\end{pmatrix}$ &
	$\begin{pmatrix}
		\times & \times & 0\\
		\times & \times & 0\\
		0 & 0 & \times
	\end{pmatrix}$ \\
	\hline
	\makecell{$\ell_L\sim([0],[0],[2])$ \\ $\nu_R\sim([1],[2])$} &
	$\begin{pmatrix}
		0 & 0 & 0\\
		0 & 0 & 0\\
		0 & 0 & \times
	\end{pmatrix}$ &
	$\begin{pmatrix}
		\times & \times & \times\\
		\times & \times & \times\\
		\times & \times & \times
	\end{pmatrix}$ &
	$\begin{pmatrix}
		\times & \times & 0\\
		\times & \times & 0\\
		0 & 0 & 0
	\end{pmatrix}$ \\
	\hline
	\makecell{$\ell_L\sim([1],[1],[0])$ \\ $\nu_R\sim([0],[1])$} &
	$\begin{pmatrix}
		\times & \times & 0\\
		\times & \times & 0\\
		0 & 0 & \times
	\end{pmatrix}$ &
	$\begin{pmatrix}
		\times & \times & 0\\
		\times & \times & 0\\
		0 & 0 & \times
	\end{pmatrix}$ &
	$\begin{pmatrix}
		\times & \times & 0\\
		\times & \times & 0\\
		0 & 0 & 0
	\end{pmatrix}$ \\
	\hline
	\makecell{$\ell_L\sim([1],[1],[0])$ \\ $\nu_R\sim([0],[2])$} &
	$\begin{pmatrix}
		0 & 0 & 0\\
		0 & 0 & 0\\
		0 & 0 & \times
	\end{pmatrix}$ &
	$\begin{pmatrix}
		\times & \times & 0\\
		\times & \times & 0\\
		0 & 0 & 0
	\end{pmatrix}$ &
	$\begin{pmatrix}
		0 & 0 & 0\\
		0 & 0 & 0\\
		0 & 0 & \times
	\end{pmatrix}$ \\
	\hline
	\makecell{$\ell_L\sim([1],[1],[0])$ \\ $\nu_R\sim([1],[2])$} &
	$\begin{pmatrix}
		\times & \times & 0\\
		\times & \times & 0\\
		0 & 0 & 0
	\end{pmatrix}$ &
	$\begin{pmatrix}
		\times & \times & 0\\
		\times & \times & 0\\
		0 & 0 & \times
	\end{pmatrix}$ &
	$\begin{pmatrix}
		\times & \times & 0\\
		\times & \times & 0\\
		0 & 0 & \times
	\end{pmatrix}$ \\
	\hline
	\makecell{$\ell_L\sim([1],[1],[2])$ \\ $\nu_R\sim([0],[1])$} &
	$\begin{pmatrix}
		\times & \times & 0\\
		\times & \times & 0\\
		0 & 0 & 0
	\end{pmatrix}$ &
	$\begin{pmatrix}
		\times & \times & 0\\
		\times & \times & 0\\
		0 & 0 & \times
	\end{pmatrix}$ &
	$\begin{pmatrix}
		\times & \times & 0\\
		\times & \times & 0\\
		0 & 0 & \times
	\end{pmatrix}$ \\
	\hline
	\makecell{$\ell_L\sim([1],[1],[2])$ \\ $\nu_R\sim([0],[2])$} &
	$\begin{pmatrix}
		0 & 0 & 0\\
		0 & 0 & 0\\
		0 & 0 & \times
	\end{pmatrix}$ &
	$\begin{pmatrix}
		\times & \times & 0\\
		\times & \times & 0\\
		0 & 0 & 0
	\end{pmatrix}$ &
	$\begin{pmatrix}
		0 & 0 & 0\\
		0 & 0 & 0\\
		0 & 0 & \times
	\end{pmatrix}$ \\
	\hline
	\makecell{$\ell_L\sim([1],[1],[2])$ \\ $\nu_R\sim([1],[2])$} &
	$\begin{pmatrix}
		\times & \times & 0\\
		\times & \times & 0\\
		0 & 0 & \times
	\end{pmatrix}$ &
	$\begin{pmatrix}
		\times & \times & 0\\
		\times & \times & 0\\
		0 & 0 & \times
	\end{pmatrix}$ &
	$\begin{pmatrix}
		\times & \times & 0\\
		\times & \times & 0\\
		0 & 0 & 0
	\end{pmatrix}$ \\
	\hline
	\makecell{$\ell_L\sim([2],[2],[0])$ \\ $\nu_R\sim([0],[1])$} &
	$\begin{pmatrix}
		0 & 0 & 0\\
		0 & 0 & 0\\
		0 & 0 & \times
	\end{pmatrix}$ &
	$\begin{pmatrix}
		\times & \times & \times\\
		\times & \times & \times\\
		\times & \times & \times
	\end{pmatrix}$ &
	$\begin{pmatrix}
		\times & \times & 0\\
		\times & \times & 0\\
		0 & 0 & 0
	\end{pmatrix}$ \\
	\hline
	\makecell{$\ell_L\sim([2],[2],[0])$ \\ $\nu_R\sim([0],[2])$} &
	$\begin{pmatrix}
		\times & \times & 0\\
		\times & \times & 0\\
		0 & 0 & \times
	\end{pmatrix}$ &
	$\begin{pmatrix}
		0 & 0 & 0\\
		0 & 0 & 0\\
		0 & 0 & 0
	\end{pmatrix}$ &
	$\begin{pmatrix}
		\times & \times & 0\\
		\times & \times & 0\\
		0 & 0 & \times
	\end{pmatrix}$ \\
	\hline
	\makecell{$\ell_L\sim([2],[2],[0])$ \\ $\nu_R\sim([1],[2])$} &
	$\begin{pmatrix}
		\times & \times & 0\\
		\times & \times & 0\\
		0 & 0 & 0
	\end{pmatrix}$ &
	$\begin{pmatrix}
		\times & \times & \times\\
		\times & \times & \times\\
		\times & \times & \times
	\end{pmatrix}$ &
	$\begin{pmatrix}
		0 & 0 & 0\\
		0 & 0 & 0\\
		0 & 0 & \times
	\end{pmatrix}$ \\
	\hline
	\makecell{$\ell_L\sim([2],[2],[1])$ \\ $\nu_R\sim([0],[1])$} &
	$\begin{pmatrix}
		0 & 0 & 0\\
		0 & 0 & 0\\
		0 & 0 & \times
	\end{pmatrix}$ &
	$\begin{pmatrix}
		\times & \times & 0\\
		\times & \times & 0\\
		0 & 0 & \times
	\end{pmatrix}$ &
	$\begin{pmatrix}
		\times & \times & 0\\
		\times & \times & 0\\
		0 & 0 & \times
	\end{pmatrix}$ \\
	\hline
	\makecell{$\ell_L\sim([2],[2],[1])$ \\ $\nu_R\sim([0],[2])$} &
	$\begin{pmatrix}
		\times & \times & 0\\
		\times & \times & 0\\
		0 & 0 & 0
	\end{pmatrix}$ &
	$\begin{pmatrix}
		0 & 0 & 0\\
		0 & 0 & 0\\
		0 & 0 & \times
	\end{pmatrix}$ &
	$\begin{pmatrix}
		\times & \times & 0\\
		\times & \times & 0\\
		0 & 0 & 0
	\end{pmatrix}$ \\
	\hline
	\makecell{$\ell_L\sim([2],[2],[1])$ \\ $\nu_R\sim([1],[2])$} &
	$\begin{pmatrix}
		\times & \times & 0\\
		\times & \times & 0\\
		0 & 0 & \times
	\end{pmatrix}$ &
	$\begin{pmatrix}
		\times & \times & 0\\
		\times & \times & 0\\
		0 & 0 & \times
	\end{pmatrix}$ &
	$\begin{pmatrix}
		0 & 0 & 0\\
		0 & 0 & 0\\
		0 & 0 & \times
	\end{pmatrix}$ \\
	\hline\midrule
\caption{\label{tab:M4-nu-seesaw}The light neutrino mass matrix $M_\nu$ in minimal seesaw from $Z_2$ gauging of $Z_4$ symmetry, where the crosses represent non-zero matrix elements. }
\end{longtable}

\subsection{ $N=5$}

The possible patterns of $Y_{\nu}$ and $M_R$ which can be derived from the $Z_2$ gauging of $Z_5$ symmetry are collected in table~\ref{tab:M5-DR-seesaw}, and the corresponding light neutrino mass matrix $M_{\nu}= -\frac{1}{2}v^2Y_{\nu}M_R^{-1}Y_{\nu}^T$ is listed in table~\ref{tab:M5-nu-seesaw}. Similar to $N=4$, here $Y_{\nu}$, $M_R$ and $M_{\nu}$ are determined up to row and column permutations.

\begin{longtable}{|c|c|c|c|c|}
	\midrule\hline
	& $Y_{\nu}\,,\,\, H\sim[0]$ & $Y_{\nu}\,,\,\, H\sim[1]$ & $Y_{\nu}\,,\,\, H\sim[2]$ & $M_R$ \\
	\hline
	\endfirsthead
	
	\hline
	  & $Y_{\nu}\,,\,\, H\sim[0]$ & $Y_{\nu}\,,\,\, H\sim[1]$ & $Y_{\nu}\,,\,\, H\sim[2]$ & $M_R$ \\
	\hline
	\endhead

	\makecell{$\ell_L\sim([0],[1],[2])$ \\ $\nu_R\sim([0],[1])$} &
	$\begin{pmatrix}
		\times & 0 \\
		0 & \times \\
		0 & 0
	\end{pmatrix}$ &
	$\begin{pmatrix}
		0 & \times\\
		\times & 0\\
		0 & \times
	\end{pmatrix}$ &
	$\begin{pmatrix}
		0 & 0 \\
		0 & \times\\
		\times & \times
	\end{pmatrix}$ &
	$\begin{pmatrix}
		\times & 0 \\
		0 & \times
	\end{pmatrix}$ \\
	\hline
	\makecell{$\ell_L\sim([0],[1],[2])$ \\ $\nu_R\sim([0],[2])$} &
	$\begin{pmatrix}
		\times & 0 \\
		0 & 0 \\
		0 & \times
	\end{pmatrix}$ &
	$\begin{pmatrix}
		0 & 0\\
		\times & \times\\
		0 & \times
	\end{pmatrix}$ &
	$\begin{pmatrix}
		0 & \times \\
		0 & \times\\
		\times & 0
	\end{pmatrix}$ &
	$\begin{pmatrix}
		\times & 0 \\
		0 & \times
	\end{pmatrix}$ \\
	\hline
	\makecell{$\ell_L\sim([0],[1],[2])$ \\ $\nu_R\sim([1],[2])$} &
	$\begin{pmatrix}
		0 & 0 \\
		\times & 0 \\
		0 & \times
	\end{pmatrix}$ &
	$\begin{pmatrix}
		\times & 0\\
		0 & \times\\
		\times & \times
	\end{pmatrix}$ &
	$\begin{pmatrix}
		0 & \times \\
		\times & \times\\
		\times & 0
	\end{pmatrix}$ &
	$\begin{pmatrix}
		\times & 0 \\
		0 & \times
	\end{pmatrix}$ \\
	\hline
	\makecell{$\ell_L\sim([0],[0],[1])$ \\ $\nu_R\sim([0],[1])$} &
	$\begin{pmatrix}
		\times & 0 \\
		\times & 0 \\
		0 & \times
	\end{pmatrix}$ &
	$\begin{pmatrix}
		0 & \times\\
		0 & \times\\
		\times & 0
	\end{pmatrix}$ &
	$\begin{pmatrix}
		0 & 0 \\
		0 & 0\\
		0 & \times
	\end{pmatrix}$ &
	$\begin{pmatrix}
		\times & 0 \\
		0 & \times
	\end{pmatrix}$ \\
	\hline
	\makecell{$\ell_L\sim([0],[0],[1])$ \\ $\nu_R\sim([0],[2])$} &
	$\begin{pmatrix}
		\times & 0 \\
		\times & 0 \\
		0 & 0
	\end{pmatrix}$ &
	$\begin{pmatrix}
		0 & 0\\
		0 & 0\\
		\times & \times
	\end{pmatrix}$ &
	$\begin{pmatrix}
		0 & \times \\
		0 & \times\\
		0 & \times
	\end{pmatrix}$ &
	$\begin{pmatrix}
		\times & 0 \\
		0 & \times
	\end{pmatrix}$ \\
	\hline
	\makecell{$\ell_L\sim([0],[0],[1])$ \\ $\nu_R\sim([1],[2])$} &
	$\begin{pmatrix}
		0 & 0 \\
		0 & 0 \\
		\times & 0
	\end{pmatrix}$ &
	$\begin{pmatrix}
		\times & 0 \\
		\times & 0 \\
		0 & \times
	\end{pmatrix}$ &
	$\begin{pmatrix}
		0 & \times \\
		0 & \times \\
		\times & \times
	\end{pmatrix}$ &
	$\begin{pmatrix}
		\times & 0 \\
		0 & \times
	\end{pmatrix}$ \\
	\hline
	\makecell{$\ell_L\sim([0],[0],[2])$ \\ $\nu_R\sim([0],[1])$} &
	$\begin{pmatrix}
		\times & 0 \\
		\times & 0 \\
		0 & 0
	\end{pmatrix}$ &
	$\begin{pmatrix}
		0 & \times\\
		0 & \times\\
		0 & \times
	\end{pmatrix}$ &
	$\begin{pmatrix}
		0 & 0 \\
		0 & 0\\
		\times & \times
	\end{pmatrix}$ &
	$\begin{pmatrix}
		\times & 0 \\
		0 & \times
	\end{pmatrix}$ \\
	\hline
	\makecell{$\ell_L\sim([0],[0],[2])$ \\ $\nu_R\sim([0],[2])$} &
	$\begin{pmatrix}
		\times & 0 \\
		\times & 0 \\
		0 & \times
	\end{pmatrix}$ &
	$\begin{pmatrix}
		0 & 0 \\
		0 & 0 \\
		0 & \times
	\end{pmatrix}$ &
	$\begin{pmatrix}
		0 & \times \\
		0 & \times \\
		\times & 0
	\end{pmatrix}$ &
	$\begin{pmatrix}
		\times & 0 \\
		0 & \times
	\end{pmatrix}$ \\
	\hline
	\makecell{$\ell_L\sim([0],[0],[2])$ \\ $\nu_R\sim([1],[2])$} &
	$\begin{pmatrix}
		0 & 0 \\
		0 & 0 \\
		0 & \times
	\end{pmatrix}$ &
	$\begin{pmatrix}
		\times & 0 \\
		\times & 0 \\
		\times & \times
	\end{pmatrix}$ &
	$\begin{pmatrix}
		0 & \times \\
		0 & \times \\
		\times & 0
	\end{pmatrix}$ &
	$\begin{pmatrix}
		\times & 0 \\
		0 & \times
	\end{pmatrix}$ \\
	\hline
	\makecell{$\ell_L\sim([1],[1],[0])$ \\ $\nu_R\sim([0],[1])$} &
	$\begin{pmatrix}
		0 & \times \\
		0 & \times \\
		\times & 0
	\end{pmatrix}$ &
	$\begin{pmatrix}
		\times & 0 \\
		\times & 0 \\
		0 & \times
	\end{pmatrix}$ &
	$\begin{pmatrix}
		0 & \times \\
		0 & \times \\
		0 & 0
	\end{pmatrix}$ &
	$\begin{pmatrix}
		\times & 0 \\
		0 & \times
	\end{pmatrix}$ \\
	\hline
	\makecell{$\ell_L\sim([1],[1],[0])$ \\ $\nu_R\sim([0],[2])$} &
	$\begin{pmatrix}
		0 & 0 \\
		0 & 0 \\
		\times & 0
	\end{pmatrix}$ &
	$\begin{pmatrix}
		\times & \times \\
		\times & \times \\
		0 & 0
	\end{pmatrix}$ &
	$\begin{pmatrix}
		0 & \times \\
		0 & \times \\
		0 & \times
	\end{pmatrix}$ &
	$\begin{pmatrix}
		\times & 0 \\
		0 & \times
	\end{pmatrix}$ \\
	\hline
	\makecell{$\ell_L\sim([1],[1],[0])$ \\ $\nu_R\sim([1],[2])$} &
	$\begin{pmatrix}
		\times & 0 \\
		\times & 0 \\
		0 & 0
	\end{pmatrix}$ &
	$\begin{pmatrix}
		0 & \times \\
		0 & \times \\
		\times & 0
	\end{pmatrix}$ &
	$\begin{pmatrix}
		\times & \times \\
		\times & \times \\
		0 & \times
	\end{pmatrix}$ &
	$\begin{pmatrix}
		\times & 0 \\
		0 & \times
	\end{pmatrix}$ \\
	\hline
	\makecell{$\ell_L\sim([1],[1],[2])$ \\ $\nu_R\sim([0],[1])$} &
	$\begin{pmatrix}
		0 & \times \\
		0 & \times \\
		0 & 0
	\end{pmatrix}$ &
	$\begin{pmatrix}
		\times & 0 \\
		\times & 0 \\
		0 & \times
	\end{pmatrix}$ &
	$\begin{pmatrix}
		0 & \times \\
		0 & \times \\
		\times & \times
	\end{pmatrix}$ &
	$\begin{pmatrix}
		\times & 0 \\
		0 & \times
	\end{pmatrix}$ \\
	\hline
	\makecell{$\ell_L\sim([1],[1],[2])$ \\ $\nu_R\sim([0],[2])$} &
	$\begin{pmatrix}
		0 & 0 \\
		0 & 0 \\
		0 & \times
	\end{pmatrix}$ &
	$\begin{pmatrix}
		\times & \times \\
		\times & \times \\
		0 & \times
	\end{pmatrix}$ &
	$\begin{pmatrix}
		0 & \times \\
		0 & \times \\
		\times & 0
	\end{pmatrix}$ &
	$\begin{pmatrix}
		\times & 0 \\
		0 & \times
	\end{pmatrix}$ \\
	\hline
	\makecell{$\ell_L\sim([1],[1],[2])$ \\ $\nu_R\sim([1],[2])$} &
	$\begin{pmatrix}
		\times & 0 \\
		\times & 0 \\
		0 & \times
	\end{pmatrix}$ &
	$\begin{pmatrix}
		0 & \times \\
		0 & \times \\
		\times & \times
	\end{pmatrix}$ &
	$\begin{pmatrix}
		\times & \times \\
		\times & \times \\
		\times & 0
	\end{pmatrix}$ &
	$\begin{pmatrix}
		\times & 0 \\
		0 & \times
	\end{pmatrix}$ \\
	\hline
	\makecell{$\ell_L\sim([2],[2],[0])$ \\ $\nu_R\sim([0],[1])$} &
	$\begin{pmatrix}
		0 & 0 \\
		0 & 0 \\
		\times & 0
	\end{pmatrix}$ &
	$\begin{pmatrix}
		0 & \times \\
		0 & \times \\
		0 & \times
	\end{pmatrix}$ &
	$\begin{pmatrix}
		\times & \times \\
		\times & \times \\
		0 & 0
	\end{pmatrix}$ &
	$\begin{pmatrix}
		\times & 0 \\
		0 & \times
	\end{pmatrix}$ \\
	\hline
	\makecell{$\ell_L\sim([2],[2],[0])$ \\ $\nu_R\sim([0],[2])$} &
	$\begin{pmatrix}
		0 & \times \\
		0 & \times \\
		\times & 0
	\end{pmatrix}$ &
	$\begin{pmatrix}
		0 & \times \\
		0 & \times \\
		0 & 0
	\end{pmatrix}$ &
	$\begin{pmatrix}
		\times & 0 \\
		\times & 0 \\
		0 & \times
	\end{pmatrix}$ &
	$\begin{pmatrix}
		\times & 0 \\
		0 & \times
	\end{pmatrix}$ \\
	\hline
	\makecell{$\ell_L\sim([2],[2],[0])$ \\ $\nu_R\sim([1],[2])$} &
	$\begin{pmatrix}
		0 & \times \\
		0 & \times \\
		0 & 0
	\end{pmatrix}$ &
	$\begin{pmatrix}
		\times & \times \\
		\times & \times \\
		\times & 0
	\end{pmatrix}$ &
	$\begin{pmatrix}
		\times & 0 \\
		\times & 0 \\
		0 & \times
	\end{pmatrix}$ &
	$\begin{pmatrix}
		\times & 0 \\
		0 & \times
	\end{pmatrix}$ \\
	\hline
	\makecell{$\ell_L\sim([2],[2],[1])$ \\ $\nu_R\sim([0],[1])$} &
	$\begin{pmatrix}
		0 & 0 \\
		0 & 0 \\
		0 & \times
	\end{pmatrix}$ &
	$\begin{pmatrix}
		0 & \times \\
		0 & \times \\
		\times & 0
	\end{pmatrix}$ &
	$\begin{pmatrix}
		\times & \times \\
		\times & \times \\
		0 & \times
	\end{pmatrix}$ &
	$\begin{pmatrix}
		\times & 0 \\
		0 & \times
	\end{pmatrix}$ \\
	\hline
	\makecell{$\ell_L\sim([2],[2],[1])$ \\ $\nu_R\sim([0],[2])$} &
	$\begin{pmatrix}
		0 & \times \\
		0 & \times \\
		0 & 0
	\end{pmatrix}$ &
	$\begin{pmatrix}
		0 & \times \\
		0 & \times \\
		\times & \times
	\end{pmatrix}$ &
	$\begin{pmatrix}
		\times & 0 \\
		\times & 0 \\
		0 & \times
	\end{pmatrix}$ &
	$\begin{pmatrix}
		\times & 0 \\
		0 & \times
	\end{pmatrix}$ \\
	\hline
	\makecell{$\ell_L\sim([2],[2],[1])$ \\ $\nu_R\sim([1],[2])$} &
	$\begin{pmatrix}
		0 & \times \\
		0 & \times \\
		\times & 0
	\end{pmatrix}$ &
	$\begin{pmatrix}
		\times & \times \\
		\times & \times \\
		0 & \times
	\end{pmatrix}$ &
	$\begin{pmatrix}
		\times & 0 \\
		\times & 0 \\
		\times & \times
	\end{pmatrix}$ &
	$\begin{pmatrix}
		\times & 0 \\
		0 & \times
	\end{pmatrix}$ \\
	\hline\midrule
\caption{\label{tab:M5-DR-seesaw}$Y_\nu$ and $M_R$ matrices in minimal seesaw from $Z_2$ gauging of $Z_4$ symmetry, where the crosses represent non-zero matrix elements.  }
\end{longtable}

\begin{longtable}{|c|c|c|c|}
	\midrule\hline
	& $M_\nu\,,\,\, H\sim[0]$ & $M_\nu\,,\,\, H\sim[1]$ & $M_\nu\,,\,\, H\sim[2]$ \\
	\hline
	\endfirsthead
	
	\hline
	& $M_\nu\,,\,\, H\sim[0]$ & $M_\nu\,,\,\, H\sim[1]$ & $M_\nu\,,\,\, H\sim[2]$ \\
	\hline
	\endhead
	
\makecell{$\ell_L\sim([0],[1],[2])$ \\ $\nu_R\sim([0],[1])$} &
	$\begin{pmatrix}
		\times & 0 & 0\\
		0 & \times & 0\\
		0 & 0 & 0
	\end{pmatrix}$ &
	$\begin{pmatrix}
		\times & 0 & \times\\
		0 & \times & 0\\
		\times & 0 & \times
	\end{pmatrix}$ &
	$\begin{pmatrix}
		0 & 0 & 0\\
		0 & \times & \times\\
		0 & \times & \times
	\end{pmatrix}$ \\
	\hline
	\makecell{$\ell_L\sim([0],[1],[2])$ \\ $\nu_R\sim([0],[2])$} &
	$\begin{pmatrix}
		\times & 0 & 0\\
		0 & 0 & 0\\
		0 & 0 & \times
	\end{pmatrix}$ &
	$\begin{pmatrix}
		0 & 0 & 0\\
		0 & \times & \times\\
		0 & \times & \times
	\end{pmatrix}$ &
	$\begin{pmatrix}
		\times & \times & 0\\
		\times & \times & 0\\
		0 & 0 & \times
	\end{pmatrix}$ \\
	\hline
	\makecell{$\ell_L\sim([0],[1],[2])$ \\ $\nu_R\sim([1],[2])$} &
	$\begin{pmatrix}
		0 & 0 & 0\\
		0 & \times & 0\\
		0 & 0 & \times
	\end{pmatrix}$ &
	$\begin{pmatrix}
		\times & 0 & \times\\
		0 & \times & \times\\
		\times & \times & \times
	\end{pmatrix}$ &
	$\begin{pmatrix}
		\times & \times & 0\\
		\times & \times & \times\\
		0 & \times & \times
	\end{pmatrix}$ \\
	\hline
	\makecell{$\ell_L\sim([0],[0],[1])$ \\ $\nu_R\sim([0],[1])$} &
	$\begin{pmatrix}
		\times & \times & 0\\
		\times & \times & 0\\
		0 & 0 & \times
	\end{pmatrix}$ &
	$\begin{pmatrix}
		\times & \times & 0\\
		\times & \times & 0\\
		0 & 0 & \times
	\end{pmatrix}$ &
	$\begin{pmatrix}
		0 & 0 & 0\\
		0 & 0 & 0\\
		0 & 0 & \times
	\end{pmatrix}$ \\
	\hline
	\makecell{$\ell_L\sim([0],[0],[1])$ \\ $\nu_R\sim([0],[2])$} &
	$\begin{pmatrix}
		\times & \times & 0\\
		\times & \times & 0\\
		0 & 0 & 0
	\end{pmatrix}$ &
	$\begin{pmatrix}
		0 & 0 & 0\\
		0 & 0 & 0\\
		0 & 0 & \times
	\end{pmatrix}$ &
	$\begin{pmatrix}
		\times & \times & \times\\
		\times & \times & \times\\
		\times & \times & \times
	\end{pmatrix}$ \\
	\hline
	\makecell{$\ell_L\sim([0],[0],[1])$ \\ $\nu_R\sim([1],[2])$} &
	$\begin{pmatrix}
		0 & 0 & 0\\
		0 & 0 & 0\\
		0 & 0 & \times
	\end{pmatrix}$ &
	$\begin{pmatrix}
		\times & \times & 0\\
		\times & \times & 0\\
		0 & 0 & \times
	\end{pmatrix}$ &
	$\begin{pmatrix}
		\times & \times & \times\\
		\times & \times & \times\\
		\times & \times & \times
	\end{pmatrix}$ \\
	\hline
	\makecell{$\ell_L\sim([0],[0],[2])$ \\ $\nu_R\sim([0],[1])$} &
	$\begin{pmatrix}
		\times & \times & 0\\
		\times & \times & 0\\
		0 & 0 & 0
	\end{pmatrix}$ &
	$\begin{pmatrix}
		\times & \times & \times\\
		\times & \times & \times\\
		\times & \times & \times
	\end{pmatrix}$ &
	$\begin{pmatrix}
		0 & 0 & 0\\
		0 & 0 & 0\\
		0 & 0 & \times
	\end{pmatrix}$ \\
	\hline
	\makecell{$\ell_L\sim([0],[0],[2])$ \\ $\nu_R\sim([0],[2])$} &
	$\begin{pmatrix}
		\times & \times & 0\\
		\times & \times & 0\\
		0 & 0 & \times
	\end{pmatrix}$ &
	$\begin{pmatrix}
		0 & 0 & 0\\
		0 & 0 & 0\\
		0 & 0 & \times
	\end{pmatrix}$ &
	$\begin{pmatrix}
		\times & \times & 0\\
		\times & \times & 0\\
		0 & 0 & \times
	\end{pmatrix}$ \\
	\hline
	\makecell{$\ell_L\sim([0],[0],[2])$ \\ $\nu_R\sim([1],[2])$} &
	$\begin{pmatrix}
		0 & 0 & 0\\
		0 & 0 & 0\\
		0 & 0 & \times
	\end{pmatrix}$ &
	$\begin{pmatrix}
		\times & \times & \times\\
		\times & \times & \times\\
		\times & \times & \times
	\end{pmatrix}$ &
	$\begin{pmatrix}
		\times & \times & 0\\
		\times & \times & 0\\
		0 & 0 & \times
	\end{pmatrix}$ \\
	\hline
	\makecell{$\ell_L\sim([1],[1],[0])$ \\ $\nu_R\sim([0],[1])$} &
	$\begin{pmatrix}
		\times & \times & 0\\
		\times & \times & 0\\
		0 & 0 & \times
	\end{pmatrix}$ &
	$\begin{pmatrix}
		\times & \times & 0\\
		\times & \times & 0\\
		0 & 0 & \times
	\end{pmatrix}$ &
	$\begin{pmatrix}
		\times & \times & 0\\
		\times & \times & 0\\
		0 & 0 & 0
	\end{pmatrix}$ \\
	\hline
	\makecell{$\ell_L\sim([1],[1],[0])$ \\ $\nu_R\sim([0],[2])$} &
	$\begin{pmatrix}
		0 & 0 & 0\\
		0 & 0 & 0\\
		0 & 0 & \times
	\end{pmatrix}$ &
	$\begin{pmatrix}
		\times & \times & 0\\
		\times & \times & 0\\
		0 & 0 & 0
	\end{pmatrix}$ &
	$\begin{pmatrix}
		\times & \times & \times\\
		\times & \times & \times\\
		\times & \times & \times
	\end{pmatrix}$ \\
	\hline
	\makecell{$\ell_L\sim([1],[1],[0])$ \\ $\nu_R\sim([1],[2])$} &
	$\begin{pmatrix}
		\times & \times & 0\\
		\times & \times & 0\\
		0 & 0 & 0
	\end{pmatrix}$ &
	$\begin{pmatrix}
		\times & \times & 0\\
		\times & \times & 0\\
		0 & 0 & \times
	\end{pmatrix}$ &
	$\begin{pmatrix}
		\times & \times & \times\\
		\times & \times & \times\\
		\times & \times & \times
	\end{pmatrix}$ \\
	\hline
	\makecell{$\ell_L\sim([1],[1],[2])$ \\ $\nu_R\sim([0],[1])$} &
	$\begin{pmatrix}
		\times & \times & 0\\
		\times & \times & 0\\
		0 & 0 & 0
	\end{pmatrix}$ &
	$\begin{pmatrix}
		\times & \times & 0\\
		\times & \times & 0\\
		0 & 0 & \times
	\end{pmatrix}$ &
	$\begin{pmatrix}
		\times & \times & \times\\
		\times & \times & \times\\
		\times & \times & \times
	\end{pmatrix}$ \\
	\hline
	\makecell{$\ell_L\sim([1],[1],[2])$ \\ $\nu_R\sim([0],[2])$} &
	$\begin{pmatrix}
		0 & 0 & 0\\
		0 & 0 & 0\\
		0 & 0 & \times
	\end{pmatrix}$ &
	$\begin{pmatrix}
		\times & \times & \times\\
		\times & \times & \times\\
		\times & \times & \times
	\end{pmatrix}$ &
	$\begin{pmatrix}
		\times & \times & 0\\
		\times & \times & 0\\
		0 & 0 & \times
	\end{pmatrix}$ \\
	\hline
	\makecell{$\ell_L\sim([1],[1],[2])$ \\ $\nu_R\sim([1],[2])$} &
	$\begin{pmatrix}
		\times & \times & 0\\
		\times & \times & 0\\
		0 & 0 & \times
	\end{pmatrix}$ &
	$\begin{pmatrix}
		\times & \times & \times\\
		\times & \times & \times\\
		\times & \times & \times
	\end{pmatrix}$ &
	$\begin{pmatrix}
		\times & \times & \times\\
		\times & \times & \times\\
		\times & \times & \times
	\end{pmatrix}$ \\
	\hline
	\makecell{$\ell_L\sim([2],[2],[0])$ \\ $\nu_R\sim([0],[1])$} &
	$\begin{pmatrix}
		0 & 0 & 0\\
		0 & 0 & 0\\
		0 & 0 & \times
	\end{pmatrix}$ &
	$\begin{pmatrix}
		\times & \times & \times\\
		\times & \times & \times\\
		\times & \times & \times
	\end{pmatrix}$ &
	$\begin{pmatrix}
		\times & \times & 0\\
		\times & \times & 0\\
		0 & 0 & 0
	\end{pmatrix}$ \\
	\hline
	\makecell{$\ell_L\sim([2],[2],[0])$ \\ $\nu_R\sim([0],[2])$} &
	$\begin{pmatrix}
		\times & \times & 0\\
		\times & \times & 0\\
		0 & 0 & \times
	\end{pmatrix}$ &
	$\begin{pmatrix}
		\times & \times & 0\\
		\times & \times & 0\\
		0 & 0 & 0
	\end{pmatrix}$ &
	$\begin{pmatrix}
		\times & \times & 0\\
		\times & \times & 0\\
		0 & 0 & \times
	\end{pmatrix}$ \\
	\hline
	\makecell{$\ell_L\sim([2],[2],[0])$ \\ $\nu_R\sim([1],[2])$} &
	$\begin{pmatrix}
		\times & \times & 0\\
		\times & \times & 0\\
		0 & 0 & 0
	\end{pmatrix}$ &
	$\begin{pmatrix}
		\times & \times & \times\\
		\times & \times & \times\\
		\times & \times & \times
	\end{pmatrix}$ &
	$\begin{pmatrix}
		\times & \times & 0\\
		\times & \times & 0\\
		0 & 0 & \times
	\end{pmatrix}$ \\
	\hline
	\makecell{$\ell_L\sim([2],[2],[1])$ \\ $\nu_R\sim([0],[1])$} &
	$\begin{pmatrix}
		0 & 0 & 0\\
		0 & 0 & 0\\
		0 & 0 & \times
	\end{pmatrix}$ &
	$\begin{pmatrix}
		\times & \times & 0\\
		\times & \times & 0\\
		0 & 0 & \times
	\end{pmatrix}$ &
	$\begin{pmatrix}
		\times & \times & \times\\
		\times & \times & \times\\
		\times & \times & \times
	\end{pmatrix}$ \\
	\hline
	\makecell{$\ell_L\sim([2],[2],[1])$ \\ $\nu_R\sim([0],[2])$} &
	$\begin{pmatrix}
		\times & \times & 0\\
		\times & \times & 0\\
		0 & 0 & 0
	\end{pmatrix}$ &
	$\begin{pmatrix}
		\times & \times & \times\\
		\times & \times & \times\\
		\times & \times & \times
	\end{pmatrix}$ &
	$\begin{pmatrix}
		\times & \times & 0\\
		\times & \times & 0\\
		0 & 0 & \times
	\end{pmatrix}$ \\
	\hline
	\makecell{$\ell_L\sim([2],[2],[1])$ \\ $\nu_R\sim([1],[2])$} &
	$\begin{pmatrix}
		\times & \times & 0\\
		\times & \times & 0\\
		0 & 0 & \times
	\end{pmatrix}$ &
	$\begin{pmatrix}
		\times & \times & \times\\
		\times & \times & \times\\
		\times & \times & \times
	\end{pmatrix}$ &
	$\begin{pmatrix}
		\times & \times & \times\\
		\times & \times & \times\\
		\times & \times & \times
	\end{pmatrix}$ \\
	\hline\midrule
\caption{\label{tab:M5-nu-seesaw}The light neutrino mass matrix $M_\nu$ in minimal seesaw from $Z_2$ gauging of $Z_4$ symmetry, where the crosses represent non-zero matrix elements. }
\end{longtable}

\section{\label{app:yetex-product} The full-rank textures of $Y_E$ and the assignments of matter fields for $N=4,5$}

From the $Z_2$ gauging of $Z_4$ symmetry, one can only obtain two textures of the charged lepton Yukawa coupling $Y_E$ with rank three up to permutations of rows and columns, as shown in Eq.~\eqref{eq:m=4-fullrankye}. The first pattern of $Y_E$ is
\begin{small}
\begin{eqnarray}
Y_E&=&\begin{pmatrix}
 \times & 0 & 0 \\
 0 & \times & 0 \\
 0 & 0 & \times
\end{pmatrix} ~\text{for}~ (\ell_L;E_R;H)\sim([0],[1],[2];[0],[1],[2];[0]) \,,\, ([0],[1],[2];[2],[1],[0];[2]) \,.
\end{eqnarray}
\end{small}
The second pattern is
\begin{eqnarray}
Y_E=\begin{pmatrix}
        \times & \times & 0 \\
        \times & \times & 0 \\
        0 & 0 & \times
    \end{pmatrix}\,,
\end{eqnarray}
which can be achieved when the lepton fields and Higgs transform in the following way
\begin{eqnarray}
(\ell_L;E_R;H)&\sim&([0],[2],[1];[1],[1],[0];[1]) \,,\, ([0],[2],[1];[1],[1],[2];[1]) \,,([0],[0],[1];[0],[0],[1];[0]) \,,\, \nonumber\\
&&\hskip-0.4in ([0],[0],[1];[1],[1],[0];[1]) \,,\, ([0],[0],[1];[1],[1],[2];[1]) \,,\, ([0],[0],[1];[2],[2],[1];[2]) \,,\nonumber\\
&&\hskip-0.4in ([0],[0],[2];[0],[0],[2];[0]) \,,\, ([0],[0],[2];[2],[2],[0];[2]) \,,\, ([1],[1],[0];[1],[1],[0];[0]) \,,\,\nonumber\\
&&\hskip-0.4in ([1],[1],[0];[0],[2],[1];[1]) \,,\, ([1],[1],[0];[0],[0],[1];[1]) \,,\, ([1],[1],[0];[2],[2],[1];[1]) \,,\, \nonumber\\
&&\hskip-0.4in ([1],[1],[0];[1],[1],[2];[2]) \,,\, ([1],[1],[2];[1],[1],[2];[0]) \,,\, ([1],[1],[2];[0],[2],[1];[1]) \,,\nonumber \\
&&\hskip-0.4in ([1],[1],[2];[0],[0],[1];[1]) \,,\, ([1],[1],[2];[2],[2],[1];[1]) \,,\, ([1],[1],[2];[1],[1],[0];[2]) \,,\nonumber\\
&&\hskip-0.4in([2],[2],[0];[2],[2],[0];[0]) \,,\, ([2],[2],[0];[0],[0],[2];[2]) \,,\, ([2],[2],[1];[2],[2],[1];[0]) \,,\,\nonumber\\
 &&\hskip-0.4in ([2],[2],[1];[1],[1],[0];[1]) \,,\, ([2],[2],[1];[1],[1],[2];[1]) \,,\, ([2],[2],[1];[0],[0],[1];[2]) \,,~~~\label{eq:m=4yeblockdiag}
\end{eqnarray}
By considering all possible assignments of fields, we find that the eight textures of $Y_E$ in Eq.~\eqref{eq:m=5-fullrankye} can be derived from the $Z_2$ gauging of $Z_5$ symmetry,
\begin{small}
\begin{eqnarray}\label{eq:diagYE-N=5}
Y_E&=&\begin{pmatrix}
\times & 0 & 0 \\
0 & \times & 0 \\
0 & 0 & \times
\end{pmatrix} ~ \text{when} ~ (\ell_L;E_R;H)\sim([0],[1],[2];[0],[1],[2];[0]) \,,\\
Y_E&=&\begin{pmatrix}
0 & \times & 0 \\
\times & 0 & \times \\
 0 & \times & \times
\end{pmatrix} ~ \text{when} ~ (\ell_L;E_R;H)\sim([0],[1],[2];[0],[1],[2];[1]) \,,\\
Y_E&=&\begin{pmatrix}
\times & \times & 0 \\
\times & \times & 0 \\
  0 & 0 & \times
\end{pmatrix} ~ \text{when} ~ (\ell_L;E_R;H)\sim([0],[2],[1];[1],[1],[0];[1]) \,,\, ([0],[0],[1];[0],[0],[1];[0]) \,, \nonumber\\
 &&\hskip1.2in  \qquad ([0],[0],[1];[1],[1],[0];[1]) \,,\, ([0],[0],[1];[1],[1],[2];[1]) \,,\,\nonumber \\
&&\hskip1.2in  \qquad ([1],[1],[0];[1],[1],[0];[0]) \,,\, ([1],[1],[0];[0],[2],[1];[1]) \,,\nonumber\\
&&\hskip1.2in  \qquad ([1],[1],[0];[0],[0],[1];[1]) \,,\, ([1],[1],[0];[2],[2],[1];[1]) \,,\,\nonumber \\
&&\hskip1.2in  \qquad ([1],[1],[2];[1],[1],[2];[0]) \,,\, ([1],[1],[2];[0],[0],[1];[1]) \,,\nonumber\\
&&\hskip1.2in  \qquad ([1],[1],[2];[2],[2],[0];[2]) \,,\label{eq:m=5yeblockdiag} \\
Y_E&=&\begin{pmatrix}
 0 & 0 & \times \\
 \times & \times & 0 \\
 \times & \times & \times
 \end{pmatrix} ~ \text{when} ~ (\ell_L;E_R;H)\sim([0],[2],[1];[1],[1],[2];[2]) \,,\, ([1],[0],[2];[1],[1],[2];[1]) \,, \\
Y_E&=&\begin{pmatrix}
0 & \times & \times \\
0 & \times & \times \\
\times & 0 & \times
\end{pmatrix} ~ \text{when} ~ (\ell_L;E_R;H)\sim([1],[1],[2];[0],[2],[1];[2]) \,,\, ([1],[1],[2];[1],[0],[2];[1]) \,, \\
Y_E&=&\begin{pmatrix}
\times & \times & 0 \\
\times & \times & 0 \\
\times & \times & \times
\end{pmatrix} ~ \text{when} ~ (\ell_L;E_R;H)\sim([0],[0],[1];[2],[2],[1];[2]) \,,\, ([1],[1],[2];[1],[1],[0];[2]) \,, \nonumber\\
 &&\hskip1.2in  \qquad ([1],[1],[2];[2],[2],[1];[1]) \,, \\
Y_E&=&\begin{pmatrix}
\times & \times & \times \\
\times & \times & \times \\
 0 & 0 & \times
\end{pmatrix} ~ \text{when} ~ (\ell_L;E_R;H)\sim([1],[1],[2];[0],[0],[2];[1]) \,,\, ([1],[1],[0];[1],[1],[2];[2]) \,, \nonumber\\
 &&\hskip1.2in  \qquad ([2],[2],[1];[1],[1],[2];[1]) \,, \\
Y_E&=&\begin{pmatrix}
\times & \times & \times \\
\times & \times & \times \\
\times & \times & 0
\end{pmatrix} ~ \text{when} ~ (\ell_L;E_R;H)\sim([1],[1],[0];[2],[2],[1];[2]) \,,\, ([1],[1],[2];[1],[1],[2];[2]) \,,\nonumber\\
 &&\hskip1.2in  \qquad ([1],[1],[2];[2],[2],[0];[1]) \,.
\end{eqnarray}
\end{small}
Notice that we can obtain the same textures by exchanging $[1]\leftrightarrow[2]$ in the field assignment, since the fusion rules are invariant under the permutation $[1]\leftrightarrow[2]$ for $N=5$.

\section{\label{app:higher-N-texproduct}The full-rank textures of $Y_E$ and $Y_\nu$ for $N=6,7$}

In this section, we report the texture zeros of the Yukawa couplings $Y_E$ and $Y_{\nu}$ which can be derived from the $Z_2$ gauging of $Z_N$ symmetries for $N=6, 7$. Generally there are plenty of assignments for the matter fields which are capable of producing the same texture, it is too lengthy to present all of them. Consequently we give a representative assignment for each texture.

From the $Z_2$ gauging of $Z_6$ symmetry, one can obtain the following textures of the charged lepton Yukawa coupling $Y_E$ with rank three up to permutations of rows and columns,
\begin{eqnarray}
Y_E&=&\begin{pmatrix}
 \times & \times & 0 \\
 \times & \times & 0 \\
 0 & 0 & \times \\
\end{pmatrix} ~ \text{when} ~ (\ell_L;E_R;H)\sim ([0],[0],[1];[0],[0],[1];[0])\,,\nonumber\\
Y_E&=&\begin{pmatrix}
 \times & 0 & 0 \\
 0 & \times & 0 \\
 0 & 0 & \times \\
\end{pmatrix} ~ \text{when} ~ (\ell_L;E_R;H)\sim ([2],[1],[0];[2],[1],[0];[0])\,,\nonumber\\
Y_E&=&\begin{pmatrix}
 \times & \times & \times \\
 \times & \times & 0 \\
 \times & \times & 0 \\
\end{pmatrix} ~ \text{when} ~ (\ell_L;E_R;H)\sim ([2],[0],[0];[1],[1],[3];[1])\,,\nonumber\\
Y_E&=&\begin{pmatrix}
 \times & \times & 0 \\
 0 & \times & 0 \\
 0 & 0 & \times \\
\end{pmatrix} ~ \text{when} ~ (\ell_L;E_R;H)\sim ([2],[0],[1];[3],[1],[0];[1])\,,\nonumber\\
Y_E&=&\begin{pmatrix}
 \times & \times & \times \\
 \times & \times & \times \\
 \times & \times & 0 \\
\end{pmatrix} ~ \text{when} ~ (\ell_L;E_R;H)\sim ([2],[2],[0];[1],[1],[3];[1])\,,\nonumber\\
Y_E&=&\begin{pmatrix}
 \times & \times & \times \\
 \times & \times & \times \\
 \times & 0 & 0 \\
\end{pmatrix} ~ \text{when} ~ (\ell_L;E_R;H)\sim ([2],[2],[0];[1],[3],[3];[1])\,,
\end{eqnarray}
which are shown in Eq.~\eqref{eq:yetex-n=6}. The above textures of $Y_E$ can also be derived from the $Z_2$ gauging of $Z_7$ symmetry with the same assignment, and one can obtain three more textures as follows,
\begin{eqnarray}
Y_E&=&\begin{pmatrix}
 \times & \times & \times \\
 \times & \times & 0 \\
 0 & 0 & \times \\
\end{pmatrix} ~ \text{when} ~ (\ell_L;E_R;H)\sim ([2],[0],[3];[1],[1],[3];[1])\,,\nonumber\\
Y_E&=&\begin{pmatrix}
 \times & \times & 0 \\
 \times & 0 & \times \\
 0 & \times & 0 \\
\end{pmatrix} ~ \text{when} ~ (\ell_L;E_R;H)\sim ([2],[3],[0];[3],[1],[2];[1])\,,\nonumber\\
Y_E&=&\begin{pmatrix}
 \times & \times & 0 \\
 \times & \times & 0 \\
 \times & 0 & \times \\
\end{pmatrix} ~ \text{when} ~ (\ell_L;E_R;H)\sim ([1],[1],[3];[2],[0],[3];[1])\,.
\end{eqnarray}

For the Dirac neutrino Yukawa matrix $Y_\nu$, from the $Z_2$ gauging of $Z_6$ symmetry, one can obtain the textures of $Y_\nu$ with rank two up to permutations of rows and columns as follows,
\begin{eqnarray}
Y_\nu&=&\begin{pmatrix}
 \times & 0 \\
 \times & 0 \\
 0 & \times \\
\end{pmatrix} ~ \text{when} ~ (\ell_L;\nu_R;H)\sim ([0],[0],[1];[0],[1];[0])\,,\nonumber\\
Y_\nu&=&\begin{pmatrix}
\times & 0 \\
 0 & \times \\
 0 & 0 \\
\end{pmatrix} ~ \text{when} ~ (\ell_L;\nu_R;H)\sim ([1],[0],[2];[1],[0];[0])\,,\nonumber\\
Y_\nu&=&\begin{pmatrix}
 \times & \times \\
 \times & 0 \\
 \times & 0 \\
\end{pmatrix} ~ \text{when} ~ (\ell_L;\nu_R;H)\sim ([2],[0],[0];[1],[3];[1])\,,\nonumber\\
Y_\nu&=&\begin{pmatrix}
 \times & \times \\
 \times & \times \\
 0 & 0 \\
\end{pmatrix} ~ \text{when} ~ (\ell_L;\nu_R;H)\sim ([1],[1],[0];[2],[0];[1])\,,\nonumber\\
Y_\nu&=&\begin{pmatrix}
 \times & \times \\
 \times & 0 \\
 0 & 0 \\
\end{pmatrix} ~ \text{when} ~ (\ell_L;\nu_R;H)\sim ([2],[0],[1];[1],[3];[1])\,,\nonumber\\
Y_\nu&=&\begin{pmatrix}
 \times & \times \\
 \times & \times \\
 \times & 0 \\
\end{pmatrix} ~ \text{when} ~ (\ell_L;\nu_R;H)\sim ([2],[2],[0];[1],[3];[1])\,,
\end{eqnarray}
which are shown in Eq.~\eqref{eq:ynutex-n=6}. Analogously the above six textures of $Y_{\nu}$ can also be obtained from the $Z_2$ gauging of $Z_7$ symmetry and the transformation rules of matter fields under non-invertible symmetry are kept intact. What's more, a new texture can be derived from the non-invertible $Z_7$ symmetry,
\begin{eqnarray}
Y_\nu&=&\begin{pmatrix}
 \times & \times \\
 \times & 0 \\
 0 & \times \\
\end{pmatrix} ~ \text{when} ~ (\ell_L;\nu_R;H)\sim ([2],[3],[0];[3],[1];[1])\,.
\end{eqnarray}

\end{appendix}

\clearpage

\providecommand{\href}[2]{#2}\begingroup\raggedright\endgroup


\begin{thebibliography}{10}

\bibitem{ParticleDataGroup:2024cfk}
{\bfseries Particle Data Group} Collaboration, S.~Navas {\em et~al.}, ``{Review
  of particle physics},''
  \href{http://dx.doi.org/10.1103/PhysRevD.110.030001}{{\em Phys. Rev. D}
  {\bfseries 110} no.~3, (2024) 030001}.

\bibitem{Capozzi:2025wyn}
F.~Capozzi, W.~Giar{\`e}, E.~Lisi, A.~Marrone, A.~Melchiorri, and A.~Palazzo,
  ``{Neutrino masses and mixing: Entering the era of subpercent precision},''
  \href{http://dx.doi.org/10.1103/PhysRevD.111.093006}{{\em Phys. Rev. D}
  {\bfseries 111} no.~9, (2025) 093006},
  \href{http://arxiv.org/abs/2503.07752}{{\ttfamily arXiv:2503.07752
  [hep-ph]}}.

\bibitem{Esteban:2024eli}
I.~Esteban, M.~C. Gonzalez-Garcia, M.~Maltoni, I.~Martinez-Soler, J.~P.
  Pinheiro, and T.~Schwetz, ``{NuFit-6.0: updated global analysis of
  three-flavor neutrino oscillations},''
  \href{http://dx.doi.org/10.1007/JHEP12(2024)216}{{\em JHEP} {\bfseries 12}
  (2024) 216}, \href{http://arxiv.org/abs/2410.05380}{{\ttfamily
  arXiv:2410.05380 [hep-ph]}}.

\bibitem{deSalas:2020pgw}
P.~F. de~Salas, D.~V. Forero, S.~Gariazzo, P.~Mart{\'\i}nez-Mirav{\'e},
  O.~Mena, C.~A. Ternes, M.~T{\'o}rtola, and J.~W.~F. Valle, ``{2020 global
  reassessment of the neutrino oscillation picture},''
  \href{http://dx.doi.org/10.1007/JHEP02(2021)071}{{\em JHEP} {\bfseries 02}
  (2021) 071}, \href{http://arxiv.org/abs/2006.11237}{{\ttfamily
  arXiv:2006.11237 [hep-ph]}}.

\bibitem{Feruglio:2019ybq}
F.~Feruglio and A.~Romanino, ``{Lepton flavor symmetries},''
  \href{http://dx.doi.org/10.1103/RevModPhys.93.015007}{{\em Rev. Mod. Phys.}
  {\bfseries 93} no.~1, (2021) 015007},
  \href{http://arxiv.org/abs/1912.06028}{{\ttfamily arXiv:1912.06028
  [hep-ph]}}.

\bibitem{Xing:2020ijf}
Z.-z. Xing, ``{Flavor structures of charged fermions and massive neutrinos},''
  \href{http://dx.doi.org/10.1016/j.physrep.2020.02.001}{{\em Phys. Rept.}
  {\bfseries 854} (2020) 1--147},
  \href{http://arxiv.org/abs/1909.09610}{{\ttfamily arXiv:1909.09610
  [hep-ph]}}.

\bibitem{Ding:2023htn}
G.-J. Ding and S.~F. King, ``{Neutrino mass and mixing with modular
  symmetry},'' \href{http://dx.doi.org/10.1088/1361-6633/ad52a3}{{\em Rept.
  Prog. Phys.} {\bfseries 87} no.~8, (2024) 084201},
  \href{http://arxiv.org/abs/2311.09282}{{\ttfamily arXiv:2311.09282
  [hep-ph]}}.

\bibitem{Ding:2024ozt}
G.-J. Ding and J.~W.~F. Valle, ``{The symmetry approach to quark and lepton
  masses and mixing},''
  \href{http://dx.doi.org/10.1016/j.physrep.2024.12.005}{{\em Phys. Rept.}
  {\bfseries 1109} (2025) 1--105},
  \href{http://arxiv.org/abs/2402.16963}{{\ttfamily arXiv:2402.16963
  [hep-ph]}}.

\bibitem{Feruglio:2025ztj}
F.~Feruglio and S.~Ramos-Sanchez, ``{Quark and lepton masses},''
  \href{http://arxiv.org/abs/2506.20755}{{\ttfamily arXiv:2506.20755
  [hep-ph]}}.

\bibitem{Weinberg:1977hb}
S.~Weinberg, ``{The Problem of Mass},''
  \href{http://dx.doi.org/10.1111/j.2164-0947.1977.tb02958.x}{{\em Trans. New
  York Acad. Sci.} {\bfseries 38} (1977) 185--201}.

\bibitem{Fritzsch:1977za}
H.~Fritzsch, ``{Calculating the Cabibbo Angle},''
  \href{http://dx.doi.org/10.1016/0370-2693(77)90408-7}{{\em Phys. Lett. B}
  {\bfseries 70} (1977) 436--440}.

\bibitem{Frampton:2002yf}
P.~H. Frampton, S.~L. Glashow, and D.~Marfatia, ``{Zeroes of the neutrino mass
  matrix},'' \href{http://dx.doi.org/10.1016/S0370-2693(02)01817-8}{{\em Phys.
  Lett. B} {\bfseries 536} (2002) 79--82},
  \href{http://arxiv.org/abs/hep-ph/0201008}{{\ttfamily arXiv:hep-ph/0201008}}.

\bibitem{Xing:2002ta}
Z.-z. Xing, ``{Texture zeros and Majorana phases of the neutrino mass
  matrix},'' \href{http://dx.doi.org/10.1016/S0370-2693(02)01354-0}{{\em Phys.
  Lett. B} {\bfseries 530} (2002) 159--166},
  \href{http://arxiv.org/abs/hep-ph/0201151}{{\ttfamily arXiv:hep-ph/0201151}}.

\bibitem{Fritzsch:1999ee}
H.~Fritzsch and Z.-z. Xing, ``{Mass and flavor mixing schemes of quarks and
  leptons},'' \href{http://dx.doi.org/10.1016/S0146-6410(00)00102-2}{{\em Prog.
  Part. Nucl. Phys.} {\bfseries 45} (2000) 1--81},
  \href{http://arxiv.org/abs/hep-ph/9912358}{{\ttfamily arXiv:hep-ph/9912358}}.

\bibitem{Gupta:2012fsl}
M.~Gupta and G.~Ahuja, ``{Flavor mixings and textures of the fermion mass
  matrices},'' \href{http://dx.doi.org/10.1142/S0217751X12300335}{{\em Int. J.
  Mod. Phys. A} {\bfseries 27} (2012) 1230033},
  \href{http://arxiv.org/abs/1302.4823}{{\ttfamily arXiv:1302.4823 [hep-ph]}}.

\bibitem{Grimus:2004hf}
W.~Grimus, A.~S. Joshipura, L.~Lavoura, and M.~Tanimoto, ``{Symmetry
  realization of texture zeros},''
  \href{http://dx.doi.org/10.1140/epjc/s2004-01896-y}{{\em Eur. Phys. J. C}
  {\bfseries 36} (2004) 227--232},
  \href{http://arxiv.org/abs/hep-ph/0405016}{{\ttfamily arXiv:hep-ph/0405016}}.

\bibitem{GonzalezFelipe:2014zjk}
R.~Gonz{\'a}lez~Felipe and H.~Ser{\^o}dio, ``{Abelian realization of
  phenomenological two-zero neutrino textures},''
  \href{http://dx.doi.org/10.1016/j.nuclphysb.2014.06.015}{{\em Nucl. Phys. B}
  {\bfseries 886} (2014) 75--92},
  \href{http://arxiv.org/abs/1405.4263}{{\ttfamily arXiv:1405.4263 [hep-ph]}}.

\bibitem{Zhang:2019ngf}
D.~Zhang, ``{A modular $A_4$ symmetry realization of two-zero textures of the
  Majorana neutrino mass matrix},''
  \href{http://dx.doi.org/10.1016/j.nuclphysb.2020.114935}{{\em Nucl. Phys. B}
  {\bfseries 952} (2020) 114935},
  \href{http://arxiv.org/abs/1910.07869}{{\ttfamily arXiv:1910.07869
  [hep-ph]}}.

\bibitem{Lu:2019vgm}
J.-N. Lu, X.-G. Liu, and G.-J. Ding, ``{Modular symmetry origin of texture
  zeros and quark lepton unification},''
  \href{http://dx.doi.org/10.1103/PhysRevD.101.115020}{{\em Phys. Rev. D}
  {\bfseries 101} no.~11, (2020) 115020},
  \href{http://arxiv.org/abs/1912.07573}{{\ttfamily arXiv:1912.07573
  [hep-ph]}}.

\bibitem{Kikuchi:2022svo}
S.~Kikuchi, T.~Kobayashi, M.~Tanimoto, and H.~Uchida, ``{Texture zeros of quark
  mass matrices at fixed point $\tau =\omega $ in modular flavor symmetry},''
  \href{http://dx.doi.org/10.1140/epjc/s10052-023-11718-1}{{\em Eur. Phys. J.
  C} {\bfseries 83} no.~7, (2023) 591},
  \href{http://arxiv.org/abs/2207.04609}{{\ttfamily arXiv:2207.04609
  [hep-ph]}}.

\bibitem{Ding:2022aoe}
G.-J. Ding, F.~R. Joaquim, and J.-N. Lu, ``{Texture-zero patterns of lepton
  mass matrices from modular symmetry},''
  \href{http://dx.doi.org/10.1007/JHEP03(2023)141}{{\em JHEP} {\bfseries 03}
  (2023) 141}, \href{http://arxiv.org/abs/2211.08136}{{\ttfamily
  arXiv:2211.08136 [hep-ph]}}.

\bibitem{Gaiotto:2014kfa}
D.~Gaiotto, A.~Kapustin, N.~Seiberg, and B.~Willett, ``{Generalized Global
  Symmetries},'' \href{http://dx.doi.org/10.1007/JHEP02(2015)172}{{\em JHEP}
  {\bfseries 02} (2015) 172}, \href{http://arxiv.org/abs/1412.5148}{{\ttfamily
  arXiv:1412.5148 [hep-th]}}.

\bibitem{Cordova:2022ruw}
C.~Cordova, T.~T. Dumitrescu, K.~Intriligator, and S.-H. Shao, ``{Snowmass
  White Paper: Generalized Symmetries in Quantum Field Theory and Beyond},'' in
  {\em {Snowmass 2021}}.
\newblock 5, 2022.
\newblock \href{http://arxiv.org/abs/2205.09545}{{\ttfamily arXiv:2205.09545
  [hep-th]}}.

\bibitem{Schafer-Nameki:2023jdn}
S.~Schafer-Nameki, ``{ICTP lectures on (non-)invertible generalized
  symmetries},'' \href{http://dx.doi.org/10.1016/j.physrep.2024.01.007}{{\em
  Phys. Rept.} {\bfseries 1063} (2024) 1--55},
  \href{http://arxiv.org/abs/2305.18296}{{\ttfamily arXiv:2305.18296
  [hep-th]}}.

\bibitem{Brennan:2023mmt}
T.~D. Brennan and S.~Hong, ``{Introduction to Generalized Global Symmetries in
  QFT and Particle Physics},''
  \href{http://arxiv.org/abs/2306.00912}{{\ttfamily arXiv:2306.00912
  [hep-ph]}}.

\bibitem{Bhardwaj:2023kri}
L.~Bhardwaj, L.~E. Bottini, L.~Fraser-Taliente, L.~Gladden, D.~S.~W. Gould,
  A.~Platschorre, and H.~Tillim, ``{Lectures on generalized symmetries},''
  \href{http://dx.doi.org/10.1016/j.physrep.2023.11.002}{{\em Phys. Rept.}
  {\bfseries 1051} (2024) 1--87},
  \href{http://arxiv.org/abs/2307.07547}{{\ttfamily arXiv:2307.07547
  [hep-th]}}.

\bibitem{Shao:2023gho}
S.-H. Shao, ``{What's Done Cannot Be Undone: TASI Lectures on Non-Invertible
  Symmetries},'' \href{http://arxiv.org/abs/2308.00747}{{\ttfamily
  arXiv:2308.00747 [hep-th]}}.

\bibitem{Kobayashi:2024cvp}
T.~Kobayashi, H.~Otsuka, and M.~Tanimoto, ``{Yukawa textures from
  non-invertible symmetries},''
  \href{http://dx.doi.org/10.1007/JHEP12(2024)117}{{\em JHEP} {\bfseries 12}
  (2024) 117}, \href{http://arxiv.org/abs/2409.05270}{{\ttfamily
  arXiv:2409.05270 [hep-ph]}}.

\bibitem{Kobayashi:2025znw}
T.~Kobayashi, Y.~Nishioka, H.~Otsuka, and M.~Tanimoto, ``{More about quark
  Yukawa textures from selection rules without group actions},''
  \href{http://dx.doi.org/10.1007/JHEP05(2025)177}{{\em JHEP} {\bfseries 05}
  (2025) 177}, \href{http://arxiv.org/abs/2503.09966}{{\ttfamily
  arXiv:2503.09966 [hep-ph]}}.

\bibitem{Kobayashi:2025ldi}
T.~Kobayashi, H.~Otsuka, M.~Tanimoto, and H.~Uchida, ``{Lepton mass textures
  from non-invertible multiplication rules},''
  \href{http://arxiv.org/abs/2505.07262}{{\ttfamily arXiv:2505.07262
  [hep-ph]}}.

\bibitem{Liang:2025dkm}
Q.~Liang and T.~T. Yanagida, ``{Non-invertible symmetry as an axion-less
  solution to the strong CP problem},''
  \href{http://dx.doi.org/10.1016/j.physletb.2025.139706}{{\em Phys. Lett. B}
  {\bfseries 868} (2025) 139706},
  \href{http://arxiv.org/abs/2505.05142}{{\ttfamily arXiv:2505.05142
  [hep-ph]}}.

\bibitem{Kobayashi:2025thd}
T.~Kobayashi, H.~Otsuka, and T.~T. Yanagida, ``{Non-invertible Symmetry as a
  Solution to the Strong CP Problem in a GUT-inspired Standard Model},''
  \href{http://arxiv.org/abs/2508.12287}{{\ttfamily arXiv:2508.12287
  [hep-ph]}}.

\bibitem{King:1999mb}
S.~F. King, ``{Large mixing angle MSW and atmospheric neutrinos from single
  right-handed neutrino dominance and U(1) family symmetry},''
  \href{http://dx.doi.org/10.1016/S0550-3213(00)00109-7}{{\em Nucl. Phys. B}
  {\bfseries 576} (2000) 85--105},
  \href{http://arxiv.org/abs/hep-ph/9912492}{{\ttfamily arXiv:hep-ph/9912492}}.

\bibitem{Frampton:2002qc}
P.~H. Frampton, S.~L. Glashow, and T.~Yanagida, ``{Cosmological sign of
  neutrino CP violation},''
  \href{http://dx.doi.org/10.1016/S0370-2693(02)02853-8}{{\em Phys. Lett. B}
  {\bfseries 548} (2002) 119--121},
  \href{http://arxiv.org/abs/hep-ph/0208157}{{\ttfamily arXiv:hep-ph/0208157}}.

\bibitem{Kageyama:2002zw}
A.~Kageyama, S.~Kaneko, N.~Shimoyama, and M.~Tanimoto, ``{Seesaw realization of
  the texture zeros in the neutrino mass matrix},''
  \href{http://dx.doi.org/10.1016/S0370-2693(02)01964-0}{{\em Phys. Lett. B}
  {\bfseries 538} (2002) 96--106},
  \href{http://arxiv.org/abs/hep-ph/0204291}{{\ttfamily arXiv:hep-ph/0204291}}.

\bibitem{Barreiros:2018ndn}
D.~M. Barreiros, R.~G. Felipe, and F.~R. Joaquim, ``{Minimal type-I seesaw
  model with maximally restricted texture zeros},''
  \href{http://dx.doi.org/10.1103/PhysRevD.97.115016}{{\em Phys. Rev. D}
  {\bfseries 97} no.~11, (2018) 115016},
  \href{http://arxiv.org/abs/1802.04563}{{\ttfamily arXiv:1802.04563
  [hep-ph]}}.

\bibitem{Bhardwaj:2022yxj}
L.~Bhardwaj, L.~E. Bottini, S.~Schafer-Nameki, and A.~Tiwari, ``{Non-invertible
  higher-categorical symmetries},''
  \href{http://dx.doi.org/10.21468/SciPostPhys.14.1.007}{{\em SciPost Phys.}
  {\bfseries 14} no.~1, (2023) 007},
  \href{http://arxiv.org/abs/2204.06564}{{\ttfamily arXiv:2204.06564
  [hep-th]}}.

\bibitem{Bartsch:2022mpm}
T.~Bartsch, M.~Bullimore, A.~E.~V. Ferrari, and J.~Pearson, ``{Non-invertible
  symmetries and higher representation theory I},''
  \href{http://dx.doi.org/10.21468/SciPostPhys.17.1.015}{{\em SciPost Phys.}
  {\bfseries 17} no.~1, (2024) 015},
  \href{http://arxiv.org/abs/2208.05993}{{\ttfamily arXiv:2208.05993
  [hep-th]}}.

\bibitem{Kobayashi:2024yqq}
T.~Kobayashi and H.~Otsuka, ``{Non-invertible flavor symmetries in magnetized
  extra dimensions},'' \href{http://dx.doi.org/10.1007/JHEP11(2024)120}{{\em
  JHEP} {\bfseries 11} (2024) 120},
  \href{http://arxiv.org/abs/2408.13984}{{\ttfamily arXiv:2408.13984
  [hep-th]}}.

\bibitem{Dong:2025jra}
J.~Dong, T.~Jeric, T.~Kobayashi, R.~Nishida, and H.~Otsuka, ``{On discrete
  gauging and non-invertible selection rules},''
  \href{http://arxiv.org/abs/2507.02375}{{\ttfamily arXiv:2507.02375
  [hep-th]}}.

\bibitem{Xing:2020ald}
Z.-z. Xing and Z.-h. Zhao, ``{The minimal seesaw and leptogenesis models},''
  \href{http://dx.doi.org/10.1088/1361-6633/abf086}{{\em Rept. Prog. Phys.}
  {\bfseries 84} no.~6, (2021) 066201},
  \href{http://arxiv.org/abs/2008.12090}{{\ttfamily arXiv:2008.12090
  [hep-ph]}}.

\bibitem{Branco:1988iq}
G.~C. Branco, L.~Lavoura, and F.~Mota, ``{Nearest Neighbor Interactions and the
  Physical Content of Fritzsch Mass Matrices},''
  \href{http://dx.doi.org/10.1103/PhysRevD.39.3443}{{\em Phys. Rev. D}
  {\bfseries 39} (1989) 3443}.

\bibitem{Branco:1994jx}
G.~C. Branco and J.~I. Silva-Marcos, ``{NonHermitian Yukawa couplings?},''
  \href{http://dx.doi.org/10.1016/0370-2693(94)91069-3}{{\em Phys. Lett. B}
  {\bfseries 331} (1994) 390--394}.

\bibitem{Branco:1999nb}
G.~C. Branco, D.~Emmanuel-Costa, and R.~Gonzalez~Felipe, ``{Texture zeros and
  weak basis transformations},''
  \href{http://dx.doi.org/10.1016/S0370-2693(00)00193-3}{{\em Phys. Lett. B}
  {\bfseries 477} (2000) 147--155},
  \href{http://arxiv.org/abs/hep-ph/9911418}{{\ttfamily arXiv:hep-ph/9911418}}.

\bibitem{Esteban_2024}
I.~Esteban, M.~C. Gonzalez-Garcia, M.~Maltoni, I.~Martinez-Soler, J.~P.
  Pinheiro, and T.~Schwetz, ``Nufit-6.0: updated global analysis of
  three-flavor neutrino oscillations,''
  \href{http://dx.doi.org/10.1007/jhep12(2024)216}{{\em Journal of High Energy
  Physics} {\bfseries 2024} no.~12, (Dec., 2024) }.
  \url{http://dx.doi.org/10.1007/JHEP12(2024)216}.

\bibitem{JUNO:2015zny}
{\bfseries JUNO} Collaboration, F.~An {\em et~al.}, ``{Neutrino Physics with
  JUNO},'' \href{http://dx.doi.org/10.1088/0954-3899/43/3/030401}{{\em J. Phys.
  G} {\bfseries 43} no.~3, (2016) 030401},
  \href{http://arxiv.org/abs/1507.05613}{{\ttfamily arXiv:1507.05613
  [physics.ins-det]}}.

\bibitem{Hyper-Kamiokande:2018ofw}
{\bfseries Hyper-Kamiokande} Collaboration, K.~Abe {\em et~al.},
  ``{Hyper-Kamiokande Design Report},''
  \href{http://arxiv.org/abs/1805.04163}{{\ttfamily arXiv:1805.04163
  [physics.ins-det]}}.

\bibitem{DUNE:2020ypp}
{\bfseries DUNE} Collaboration, B.~Abi {\em et~al.}, ``{Deep Underground
  Neutrino Experiment (DUNE), Far Detector Technical Design Report, Volume II:
  DUNE Physics},'' \href{http://arxiv.org/abs/2002.03005}{{\ttfamily
  arXiv:2002.03005 [hep-ex]}}.

\end{thebibliography}
\end{document}